\documentclass[preprint,ij4uq,inline]{ij4uq}      

\usepackage[utf8]{inputenc} 
\usepackage{mathtools}
\usepackage{bbm}
\usepackage{csl}
\usepackage{cleveref}
\usepackage[hang]{footmisc}
\def\*#1{\boldsymbol{\mathbf{#1}}}
\def\!#1{\mathcal{#1}}

\newif\ifreport

\newcommand{\titlestring}{A copula-based rank histogram ensemble filter}
\newcommand{\titlestringshort}{A copula-based rank histogram ensemble filter}
\newcommand{\authorstring}{Amit N Subrahmanya, Julie Bessac, Andrey A. Popov, Adrian Sandu}
\newcommand{\authorstringshort}{A. N. Subrahmanya, J. Bessac, A. A. Popov, \& A. Sandu}
\newcommand{\emailstring}{amitns@vt.edu, julie.bessac@nrel.gov, apopov@hawaii.edu, sandu@vt.edu}

\newcommand{\x}{\*{x}}

\newcommand{\xe}{\*{x}^{[e]}}
\newcommand{\xae}{\*{x}^{\mathrm{a}[e]}}

\newcommand{\xxe}{x^{[e]}}
\newcommand{\xxa}{x^\mathrm{a}}
\newcommand{\xxae}{x^{\mathrm{a}[e]}}

\newcommand{\zze}{z^{[e]}}

\newcommand{\zzae}{z^{\mathrm{a}[e]}}

\newcommand{\ue}{\*{u}^{[e]}}
\newcommand{\uue}{u^{[e]}}

\newcommand{\xk}{\*{x}^k}
\newcommand{\xtrk}{\*{x}^{\mathrm{tr},k}}
\newcommand{\xak}{\*{x}^{\mathrm{a},k}}

\newcommand{\z}{\*{z}}
\newcommand{\za}{\*{z}^{\mathrm{a}}}
\newcommand{\ztrk}{\*{z}^{\mathrm{tr},k}}

\newcommand{\ze}{\*{z}^{[e]}}
\newcommand{\zae}{\*{z}^{\mathrm{a}[e]}}

\newcommand{\y}{\*{y}}
\newcommand{\yk}{\*{y}^k}

\newcommand{\obserr}{\boldsymbol{\eta}}

\newcommand{\Hn}{\!H}
\newcommand{\Mn}{\!M}

\newcommand{\dif}{\mathrm{d}}

\newcommand{\pr}{p}

\newcommand{\pro}{{p^{\mathrm{o}}}}

\newcommand{\prok}{p^{k,\mathrm{o}}}

\newcommand{\prc}{P}

\newcommand{\cpj}{C}
\newcommand{\cpd}{c}

\newcommand{\Rspace}{\mathbb{R}}

\newcommand{\nens}{n_\mathrm{ens}}
\newcommand{\nstate}{n_\mathrm{st}}
\newcommand{\nk}{n_\mathrm{k}}

\newcommand{\nobs}{n_\mathrm{obs}}

\newtheorem{example}{Example}[section]

\renewcommand{\mid}{\mkern1mu | \mkern2mu}

\reporttrue

\fancypagestyle{plain}{%
	\fancyhf{}
	\ifreport
	\fancyhead[R]{\small {\it Preprint}}
	\else
	\fancyhead[R]{\small {\it International Journal for Uncertainty Quantification}, x(x): \thepage--\pageref{LastPage} (\myyear\today)}
	\fi
	\fancyfoot[R]{\small\bf\thepage }
	\fancyfoot[L]{\fottitle}
}

\renewcommand{\myyear}{2025}
\renewcommand{\today}{}

\begin{document}
	
	\ifreport
	\csltitle{\titlestring}
	\cslauthor{\authorstring}
	\cslyear{25}
	\cslreportnumber{2}
	\cslemail{\emailstring}
	\csltitlepage
	\fi
	
	\ifreport
	\volume{}
	\else
	\volume{Volume x, Issue x, \myyear\today}
	\fi
	
	\title{\titlestring}
	\titlehead{\titlestringshort}
	\authorhead{\authorstringshort}
	\corrauthor[1]{Amit N. Subrahmanya}
	\author[2]{Julie Bessac}
	\author[3]{Andrey A. Popov}
	\author[1]{Adrian Sandu}
	\corremail{amitns@vt.edu}
	\corraddress{}
	\address[1]{Computational Science Laboratory, Department of Computer Science, Virginia Tech, Blacksburg, Virginia, 24061}
	\address[2]{National Renewable Energy Laboratory, Golden, Colorado, 80401}
	\address[3]{Department of Information \& Computer Sciences, University of Hawai'i at Manoa, Honolulu, Hawai'i, 96822}

	
	\dataO{mm/dd/yyyy}
	\dataF{mm/dd/yyyy}
	
	\abstract{
		Serial ensemble filters implement triangular probability transport maps to reduce high-dimensional inference problems to sequences of state-by-state univariate inference problems.
		The univariate inference problems are solved by sampling posterior probability densities obtained by combining constructed prior densities with observational likelihoods according to Bayes' rule.  
		Many serial filters in the literature focus on representing the marginal posterior densities of each state. However, rigorously capturing the conditional dependencies between the different univariate inferences is crucial to correctly sampling multidimensional posteriors. 
		This work proposes a new serial ensemble filter, called the copula rank histogram filter (CoRHF), that seeks to capture the conditional dependency structure between variables via empirical copula estimates; these estimates are used to rigorously implement the triangular (state-by-state univariate) Bayesian inference.
		The success of the CoRHF is demonstrated on two-dimensional examples and the Lorenz '63 problem.
		A practical extension to the high-dimensional setting is developed by localizing the empirical copula estimation, and is demonstrated on the Lorenz '96 problem.
	}
	
	\keywords{Bayesian Inference, Data Assimilation, Particle Filters, Uncertainty Quantification}
	
	\maketitle
	
	\section{Introduction}
	
	Data assimilation seeks to estimate the unknown true state of a dynamical system by combining an uncertain initial state estimate, an approximate computational model of the said system, and noisy measurements of the unknown true state~\cite{Asch_2016_book,Reich_2015_book,Evensen_2022_book}.
	Originating in numerical weather prediction, data assimilation has successfully been used in many areas such as ocean modeling, geophysics, economics, digital twins, etc.
	A popular family of data assimilation algorithms includes the ensemble Kalman filter and its innumerable variants~\cite{Evensen_1994_EnKF,Burgers_1998_EnKF,Houtekamer_1998_EnKF,Bishop_2001_ETKF,Evensen_2003_EnKF,Anderson_2001_EAKF,Anderson_2003_Local,Sakov_2008_DEnKF} which are optimal when the model dynamics and observation operator are linear, and the state uncertainty and observational likelihood are Gaussian. 
	While these filters show acceptable performance in moderately non-linear, non-Gaussian settings, it is of utmost interest to investigate filters that are close to optimal for highly non-linear, non-Gaussian, multimodal settings. 
	Another popular family of algorithms includes particle filters~\cite{vanLeeuwen_2009_PFreview,vanLeeuwen_2019_PFreview}, which make no restrictive Gaussianity assumptions, and should perform optimal inference under any setting. 
	However, in high dimensions, an infeasibly large number of samples (or particles) is required to prevent filter degeneracy.
	Newer particle filters that modify the states in addition to modifying the weights~\cite{Reich_2013_ETPF,Subrahmanya_2023_Ensemble,Popov_2024_EnGMF,Hu_2024_PFF} have strong potential for high-dimensional filtering, but further research is needed to make these filters applicable to operational problems.

	To efficiently tackle data assimilation problems in high dimensions, \cite{Anderson_2003_Local} proposed a serial filtering approach, where high-dimensional inference problems are solved through sequences of state-by-state univariate inference problems.
	The system is written in a joint state and observation space; the observation increments are calculated first based on a univariate ensemble filter with some approximation to the observational likelihood (a Gaussian approximation results in a scheme very similar to the ensemble Kalman filter); the state increments are calculated next based on a linear regression of the observation increments.
	The rank histogram filter (RHF)~\cite{Anderson_2010_Rhf} extends the serial approach to non-Gaussian systems via deterministic sampling of the observed states, where one reconstructs each univariate prior based on the order statistics (rank histogram) of the univariate ensemble; increments of the unobserved states are obtained by linear regression.
	\cite{Metref_2014_MRHF} extended the RHF to a multivariate framework by connecting serial univariate inferences to the Knothe-Rosenblatt rearrangement and calculating multidimensional regression covariance based on spatial locality.
	The marginal adjustment rank histogram filter (MARHF)~\cite{Anderson_2020_Marhf} applies a standard filter, such as the EnKF, to obtain a standard analysis, and also applies the RHF independently to every observed state variable. 
	Each RHF posterior value for a state variable is then assigned to ensemble members such that they have the same rank statistics as the marginal from the standard method.
	The quantile-conserving ensemble filter framework (QCEFF)~\cite{Anderson_2022_Qceff1,Anderson_2023_Qceff2,Anderson_2024_Qceff3} further generalizes the RHF by using Gaussian anamorphosis to perform regression in the standard probit space, rather than in primitive state space, thus allowing for sampling from weakly non-Gaussian priors.
	Gaussian anamorphosis for data assimilation is not new and has been previously explored by \cite{Simon_2012_Anamorphosis}, and \cite{Amezcua_2014_Anamorphosis}.
	\cite{Grooms_2022_Nonlinear} extends the serial filtering framework to non-linear observations.
	
	The serial filters can be interpreted in terms of measure transportation theory as triangular maps~\cite{Santambrogio_2015_OT,Villani_2009_OT}.
	Triangular maps~\cite{Bogachev_2005_Triangle} are named as such because the transportation map for any scalar state variable depends on only the previous (ordered lexicographically) state variables.
	In the multivariate rank histogram filter (MRHF), \cite{Metref_2014_MRHF} connects serial univariate inference to the Knothe-Rosenblatt rearrangement.
	\cite{Spantini_2018_Coupling,Spantini_2022_Coupling} explore nonlinear triangular maps using the Knothe-Rosenblatt rearrangement to sample from Bayesian posteriors in a univariate fashion. 
	Triangular map ensemble smoothing is done in \cite{Ramgraber_2023_ETS1,Ramgraber_2023_ETS2}. 
	
	\subsection{Contributions}
	The RHF, QCEFF, and MRHF do not accurately sample the posterior in the case of strongly non-Gaussian, multimodal densities, because they do not fully account for the conditional dependencies between the observations and hidden states.
	This work brings the following novel contributions.
	\begin{enumerate*}[label=(\roman*)]
		\item The rank histogram filtering approaches (RHF, MARHF, QCEFF) are rigorously extended to highly nonlinear, non-Gaussian, multimodal distributions by employing conditional copula densities to account for the coupling between the observed and the hidden states. 
		Algorithmically, the copulas bring additional terms to the univariate inferences that behave like the observational likelihoods.
		The new approach is named the copula rank histogram filter (CoRHF).
		\item The work develops a methodology to accurately estimate conditional copula densities via kernel density estimation in the multivariate uniform marginal space. 
		\item A spatial localization framework is proposed to estimate conditional copula densities in high-dimensional problems. 
		\item The accuracy gains of CoRHF versus traditional filters are demonstrated on standard test problems.
	\end{enumerate*}
	
	\subsection{Organization}
	The remainder of the paper is organized as follows. 
	Immediately following the organization, we discuss the notation we use throughout this manuscript.
	\Cref{sec:background} reviews data assimilation, serial filtering, and copula theory. 
	The copula rank histogram filter (CoRHF) is derived in \Cref{sec:crhf}, and practical aspects of conditional copula density estimation and extensions to high dimensions via localization are discussed in \Cref{sec:crhf-implementation}.
	\Cref{sec:expt} demonstrates the strength of CoRHF on the Lorenz '63 and Lorenz '96 test problems.   
	Finally, \Cref{sec:conc} draws conclusions and points to future work directions.
	
	\subsection{Notation}
	An $n$-dimensional random variable in the Euclidean real space is represented by $\x \in \Rspace^{n}$, whose elements are given by $\x = [x_1,\dots,x_n]$. 
	A subrange of random variables is denoted with the respective indices in the subscript, for example, for $j \ge i$ we have $\x_{i:j} = [x_i,\dots,x_j]$.
	With a slight abuse of notation, here $\x$ refers to both the random variable and its realization. 
	The probability density function (PDF) and the cumulative distribution function (CDF) of the random variable $\x$ are denoted by $\pr(\x)$ and $\prc_{\x}(\x)$, respectively.
	Similarly, the PDF and CDF of an element of the random variable $x_i$ are $\pr(x_i)$ and $\prc_{x_i}(x_i)$, respectively.
	The ``marginal uniform'' random variable $\*{u} \in [0, 1]^n$ of $\x$ is obtained by applying the probability integral transform to each element of $\x$ as $\*u = [u_1,\dots,u_n] = [\prc_{x_1}(x_1),\dots,\prc_{x_n}(x_n)]$.
	In the univariate case, as the cumulative distribution is a monotonically non-decreasing function, it is assumed to have a pseudo-inverse defined as $\prc_{x}^{-1}(u) = \inf \{ u \in \Rspace, \prc_x(x) \geq u \}$.
	While the pseudo-inverse is required for theoretical discussion, all the cumulative densities arising in this work are strictly increasing and hence invertible.
	The $e$-th sample in an ensemble of $\nens$ samples is denoted by $\xe$ for $e \in \{1 , 2, \dots, \nens \}$.
	%
	
	\section{Background}
	\label{sec:background}
	
	Data assimilation aims to combine observations of an unknown dynamical system with numerical simulations of the said system. 
	Due to the uncertainties involved, the states of the numerical simulation are modeled as a random vector evolving in time as a Markov process.
	The state of the numerical simulation at time $t^k$ is denoted by $\xk \in \Rspace^{\nstate}$, and is evolved in time by the (non-)linear dynamical model $\Mn$:
	\begin{equation}
		\label{eq:model}
		\x^{k+1} = \Mn^k(\xk), \quad \Mn : \Rspace^{\nstate} \to \Rspace^{\nstate}.
	\end{equation}
	This paper assumes a fully deterministic model dynamics without model error.
	%
	
	Measurements taken at discrete time instance $t^k$ are random variables $\yk \in \Rspace^{\nobs}$ which are related to the state as follows:
	\begin{equation}
		\label{eq:measurement}
		\yk = \Hn^{k}(\xtrk) + \obserr^k, \quad {\Hn}^k:\Rspace^{\nstate} \to \Rspace^{\nobs}, \quad \obserr^k \sim \prok \in \Rspace^{\nobs},
	\end{equation}
	where $\Hn^k$ is the (non-)linear observation operator, $\xtrk$ is the latent true state and $\obserr^k$ is the observation error sampled from the density $\prok$.
	The observation errors $\obserr^k$ are assumed to be independent and identically distributed across time.
	In many problems of interest, the observation operator and the observation error distribution do not change with time.
	We denote by $ \y^{1 : k} \coloneqq [\y^1,\dots,\y^k]$ the sequence of measurements from time $t^1$ up to, and including, time $t^k$.
	
	The model observable variables \cite{Anderson_2003_Local,Anderson_2001_EAKF,Grooms_2022_Nonlinear} are the model prediction of the measured values, i.e., the model state projected onto the observation space (as in \cref{eq:measurement}):
	\begin{equation}
		\label{eq:observables}
		\z^k \coloneqq \Hn^k(\x^k) \in \Rspace^{\nobs} \quad \Rightarrow \quad
		\yk = \ztrk + \obserr^k.
	\end{equation}
	
	The goal of filtering is to estimate the following sequence of conditional PDFs $\pr(\xk \mid \y^{1 : k})$ at every $t^k$, $k \ge 1$. Using Bayes' rule, at each $t^k$, one combines a prior (or forecast) density $\pr\left(\xk \mid  \y^{1 : k - 1}\right)$ with the observational likelihood $\pr\left(\yk \mid \xk, \y^{1 : k - 1}\right)$ to obtain the posterior (or analysis) $\pr\left(\xk \mid \y^{1 : k}\right)$
	\begin{equation}
		\label{eqn:bayesian-posterior}
		\begin{split}
			\pr(\xk \mid \y^{1 : k}) 
			&= \frac{\pr\left(\yk \mid \xk, \y^{1 : k - 1}\right) \int \pr\left(\xk \mid \x^{k - 1}\right) \, \pr\left(\x^{k - 1} \mid \y^{1 : k - 1}\right) \, \dif\x^{k - 1}}{\int \pr\left(\yk \mid \xk\right) \pr\left( \xk \mid  \y^{1 : k - 1}\right)\, \dif\xk} \\
			& = \frac{\pr\left(\yk \mid \xk, \y^{1 : k - 1}\right) \pr \left(\xk \mid \y^{1 : k - 1} \right)}{\pr\left(\yk \mid \y^{1 : k - 1}\right)}.
		\end{split}
	\end{equation}
	As this work considers filtering only (i.e., assimilating data from each time instance, then propagating the state using the model to the next time instance), we will use the following notation for simplicity.
	The prior density is denoted by $\pr(\x) \equiv \pr\left(\xk \mid  \y^{1 : k - 1}\right)$, the observational likelihood as $\pr(\y \mid \x) \equiv \pr\left(\yk \mid \xk, \y^{1 : k - 1}\right)$ and the posterior density as $\pr(\x \mid \y) \equiv \pr\left(\x^{k} \mid \y^{1 : k}\right)$.  
	This Bayesian posterior (\cref{eqn:bayesian-posterior}) in the simpler notation reads: 
	\begin{equation}
		\pr(\x \mid \y) = \frac{\pr(\y \mid \x)}{\pr(\y)}\, \pr(\x).
	\end{equation}
	As in \cite{Anderson_2003_Local,Anderson_2001_EAKF,Grooms_2022_Nonlinear}, one can write the Bayes' rule in the joint state-observation space $[\z,\x]$ with $\z \coloneqq \Hn(\x)$ as follows:
	\begin{equation}
		\label{eq:bayes_cond}
		\begin{split}
			\pr(\z, \x \mid \y) = \frac{\pr(\y \mid \z, \x)}{\pr(\y)} \, \pr(\z, \x) = \frac{\pr(\y \mid \z, \x)}{\pr(\y)} \, \pr(\x \mid \z) \, \pr(\z)\\= \left(\frac{\pr(\y \mid \z)\, \pr(\z)}{\pr(\y)} \right)\, \pr(\x \mid \z) = \pr(\z \mid \y)\, \pr(\x \mid \z).
		\end{split}
	\end{equation}
	Data assimilation assumes that $\pro(\obserr) = \pr(\y \mid \x) = \pr(\y - \Hn(\x)) \equiv \pr(\y - \z)$ \cite{Evensen_2022_book}, rendering $\y$ conditionally independent of $\x$ given $\z$, i.e., $\pr(\y \mid \z, \x) = \pr(\y \mid \z)$.
	%
	%
	It is infeasible to work directly with probability densities in high dimensions. 
	A practical Monte-Carlo approach represents probability distributions by samples (particles or ensemble members) and leads to ensemble-based filters. 
	The filtering goal is to obtain posterior samples (particles) $\xae \sim \pr(\x \mid \y)$ for all $e \in \{ 1, \dots, \nens\}$ given prior samples (particles) $\xe \sim \pr(\x)$, observation $\y$, and observational likelihood $\pr(\y \mid \x)$.
	Among the many kinds of ensemble and particle filters available, the following discussion targets serial ensemble filters.

	\subsection{Rank histogram-based filters}
	\label{ssec:RHF}
	We start with discussing \Cref{alg:rhf-like} that serves as a general framework to describe serial filtering.
	\begin{algorithm}
		\caption{General univariate inference for an arbitrary scaling function.}
		\label{alg:rhf-like}
		\KwIn{Samples : $\chi^{[e]} \in \Rspace$ for all $e \in \{1, \dots,\nens\}$, and scaling function : $g(\chi) : \Rspace \to \Rspace$.}
		Order the samples such that $\chi^{[i]} < \chi^{[j]}$ for $1 \le i < j \le \nens$.\\
		Construct an approximate density estimate $\pr(\chi)$ which is piecewise continuous in $\Rspace$.\\
		Evaluate $g(\chi^{[e]})$ for all $e \in \{1, \dots,\nens\}$.\\
		Construct a piecewise continuous $g(\chi)$ using $g(\chi^{[e]})$, which is defined wherever $\pr(\chi) > 0$.
		In this work,
		\begin{equation}
			\label{eq:piecewise-g}
			g(\chi) = 
			\begin{cases} 
				\frac{1}{2}\, g(\chi^{[1]}), & \chi \le  \chi^{[1]}, \\ 
				\frac{1}{2}\, g(\chi^{[e]}) + \frac{1}{2}\, g(\chi^{[e + 1]}), &  \chi^{[e]} < \chi \le \chi^{[e + 1]},~~ 1 \le e \le \nens-1, \\
				\frac{1}{2}\, g(\chi^{[\nens]}), & \chi > \chi^{[\nens]}.
			\end{cases}
		\end{equation}\\
		Calculate the scaled density as $\widetilde{p}(\chi) = \frac{1}{C}\,g(\chi)p(\chi)$ where $C = \int g(\chi)p(\chi)$.\\
		Compute the scaled cumulative distribution function $\widetilde{\prc}_{\chi}(\chi)$.\\
		\KwRet{$\widetilde{\prc}_{\chi}(\chi)$.}
	\end{algorithm}
	\Cref{alg:rhf-like} multiplies the density estimate of scalar samples with a scaling function and returns a scaled cumulative distribution function.
	Next, samples from the scaled distribution $\widetilde{\chi}^{[e]} \sim \widetilde{\pr}_{\chi}(\cdot)$ are obtained by the integral transform of samples from a uniform random variable $\widetilde{\chi}^{[e]} = \widetilde{\prc}_{\chi}^{-1}(\uue)$, where uniform random variable samples are drawn using one of the following three approaches:
	\begin{subequations}
		\begin{alignat}{2}
			\label{eq:samp-sto}
			&\text{(stochastic)  } &&\uue \sim \operatorname{Uniform}(0, 1); \\
			\label{eq:samp-qsto}
			&\text{(quantile stochastic)  } &&\uue \sim  \operatorname{Uniform}\left(\left\{ \frac{1}{\nens + 1}, \frac{2}{\nens + 1}, \dots, \frac{\nens}{\nens + 1} \right\}\right); \\
			\label{eq:samp-qdet}
			&\text{(quantile deterministic)  } &&\uue = \frac{e}{\nens + 1}, \quad e=1,\dots,\nens.
		\end{alignat}
	\end{subequations}
	The quantile stochastic sampling can be done with or without replacement.
	The implementation details concerning the construction of the density estimate and scaling function approximations in \Cref{alg:rhf-like} are described below with the help of \Cref{fig:rhfs}.
	In this work, the function approximations closely follow the ideas from the RHF and QCEFF. 
	
	%
	\subsubsection{RHF}
	\label{sssec:RHF}
\begin{figure*}[!ht]    
		\centering
		\includegraphics[width=\linewidth]{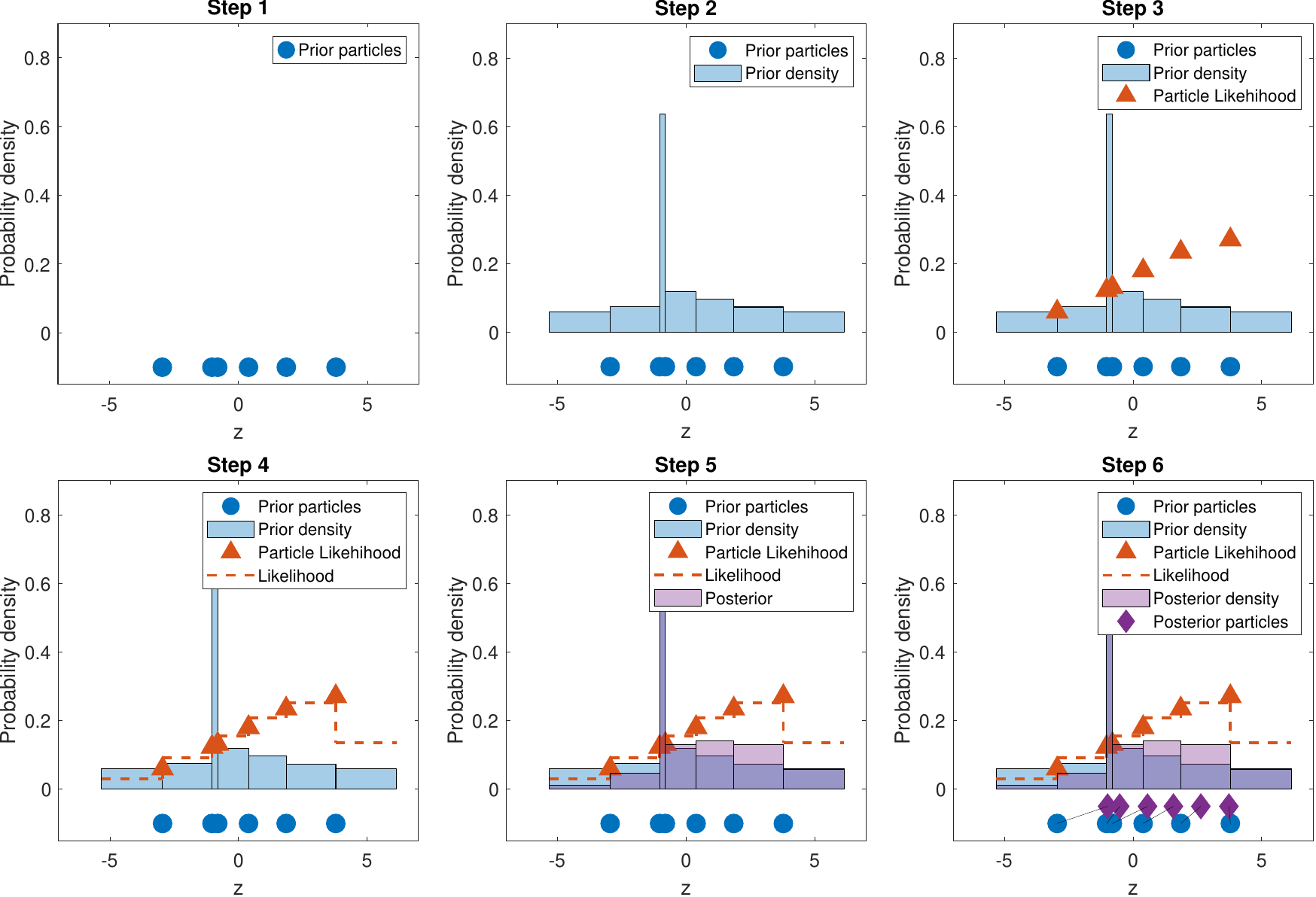}
		\caption{Univariate Bayesian inference using the rank histogram filter.}
		\label{fig:rhfs}
	\end{figure*}

	%
	For a univariate inference problem, the posterior cumulative distribution $\prc_{z \mid y}$ of the rank histogram filter~\cite{Anderson_2010_Rhf} is obtained by calling \Cref{alg:rhf-like} with the input samples being the prior particles $\zze$, and the scaling function being the observational likelihood $\pr(y \mid z)$.
	The resulting CDF of \Cref{alg:rhf-like}---the posterior distribution---is sampled according to \cref{eq:samp-qdet} to obtain the analysis particles $\zzae$:
	\begin{enumerate}[label=(\roman*)]
		\item For step 1 in \Cref{alg:rhf-like}, consider a set of $\nens$ prior particles ordered by value such that $z^{[i]} < z^{[j]}$ for $1 \le i < j \le \nens$ ($\nens = 6$ in step 1 of \Cref{fig:rhfs}). 
		\item For step 2 in \Cref{alg:rhf-like}, construct a piecewise continuous probability density $\pr_z$ to represent the given prior particles, as shown in step 2 of \Cref{fig:rhfs}. 
		In this work, a piecewise constant density is assumed such that the cumulative probability distribution between any two consecutive particles is equal, giving the following marginal uniform for each particle:
		\begin{equation}
			\uue = \prc_{z}(\zze) = \frac{e}{\nens + 1} \text{ for all } e \in \{ 1, \dots, \nens\}.
		\end{equation} 
		The tails are also constructed to have an area of $\frac{1}{\nens + 1}$.
		For simplicity, this work uses flat tails that terminate after a certain length.
		Of course, any approximation from \cite{Anderson_2022_Qceff1,Grooms_2024_QCEFF} can be used to construct the prior density.
		\item In step 3 of \Cref{alg:rhf-like}, the observational likelihood $\pr_{y \mid z}(y \mid \zze)$ is evaluated for each particle as depicted in step 3 of \Cref{fig:rhfs}.
		\item Next, construct an approximate piecewise continuous likelihood function as shown in step 4 of \Cref{fig:rhfs} and described in \Cref{alg:rhf-like}. 
		The likelihood in the interval between two consecutive particles is piecewise constant and equal to the average of the likelihoods of the said particles, i.e., 
		\begin{equation}
			\label{eq:obs-lik}
			\pr_{y \mid z}(y \mid z) = 
			\begin{cases} 
				\frac{1}{2}\, \pr_{y \mid z}(y \mid z^{[1]}), & z \le  z^{[1]}, \\ 
				\frac{1}{2}\, \pr_{y \mid z}(y \mid \zze) + \frac{1}{2}\, \pr_{y \mid z}(y \mid z^{[e + 1]}), & \zze < z \le z^{[e + 1]},~~ 1 \le e \le \nens-1, \\
				\frac{1}{2}\, \pr_{y \mid z}(y \mid z^{[\nens]}), & z > z^{[\nens]}.
			\end{cases}
		\end{equation}
		The above likelihood is constructed to reduce the posterior calculation to a multiplication of $\nens + 1$ scalars.
		However, one may certainly use the exact continuous likelihood for this.
		\item As in step 5 of both \Cref{alg:rhf-like} and \Cref{fig:rhfs}, the posterior is now estimated using Bayes' rule by multiplying the prior density with the observational likelihood, $\pr_{z \mid y}(z \mid y) \propto \pr_{y \mid z}(y \mid z)\, \pr_z(z)$ for all $z$, and rescaling it such that $\int_{-\infty}^{\infty} \pr(z \mid y) dz = 1$.
		\item Finally, the posterior $\pr_{z \mid y}$ is sampled as in \cref{eq:samp-qdet} using the inverse cumulative density function to give the posterior particles $\zzae = \prc_{z \mid y}^{-1} \left(\frac{e}{\nens + 1}\right)$ for all $e \in \{1,\dots, \nens \}$, as shown in step 6 of \Cref{fig:rhfs}.
	\end{enumerate}

	In a multivariate case, one can apply RHF independently on each observable (component of $\z$) as
	\begin{equation}
		\label{eqn:formula-simple}
		\zzae_i = \prc_{z_i \mid y_i}^{-1}\left( {\frac{e}{\nens + 1}} \right),
		\quad i = 1, \dots, \nstate, \quad e \in \{ 1, \dots, \nens \}.
	\end{equation}
	\Cref{eqn:formula-simple} assumes that the posterior $\pr(\z \mid \y)$ from \cref{eq:bayes_cond} can be factorized as:  
	\begin{equation}
		\label{eq:obs-independence}
		\begin{split}
			\pr(\z \mid \y) &= \frac{\pr(\y \mid \z)\, \pr(\z)}{\pr(\y)} = \frac{\pr(y_1 \mid \z)\, \pr(z_1)}{\pr(y_1)} \prod_{i = 2}^{\nobs} \frac{\pr(y_i \mid \y_{1:i-1}, \z)\, \pr(z_i \mid \z_{1:i-1})}{\pr(y_i \mid \y_{1:i-1})}\\
			&= \frac{\pr(y_1 \mid z_1)\, \pr(z_1)}{\pr(y_1)} \prod_{i = 2}^{\nobs} \frac{\pr(y_i \mid z_i)\, \pr(z_i \mid \z_{1:i-1})}{\pr(y_i \mid \y_{1:i-1})} \\
			&\approx \frac{\pr(y_1 \mid z_1)\, \pr(z_1)}{\pr(y_1)} \prod_{i = 2}^{\nobs} \frac{\pr(y_i \mid z_i)\, \pr(z_i)}{\pr(y_i)} = \prod_{i = 1}^{\nobs} \pr(z_i \mid y_i)
		\end{split}	
	\end{equation}
	Reading left to right, top to bottom in \cref{eq:obs-independence}, 
	\begin{enumerate*}[label=(\roman*)]
		\item the first equality comes from Bayes' rule;
		\item the second equality comes from the chain rule for joint probability applied as $\pr(\y) = \pr(y_1)\prod_{i = 2}^{\nobs} \pr(y_i \mid \y_{1:i-1})$, $\; \pr(\z) = \pr(z_1)\prod_{i = 2}^{\nobs} \pr(z_i \mid \z_{1:i-1})$, and $\pr(\y \mid \z) = \pr(y_1 \mid \z)\prod_{i = 2}^{\nobs} \pr(y_i \mid \y_{1:i-1} \z)$ ;
		\item the third equality comes from the conditional independence of $y_i$ from $\y_{1:i-1}$ and $z_j$ for $j \in \{1,\dots,\nobs \},\; j \neq i$ given $z_i$ because $z_i$ is an estimate the observed data $y_i$;
		\item the approximation assumes that element of $y_i$ is independent of $y_j$ and $z_i$ is independent of $z_j$ where $j \in \{1,\dots,\nobs \},\; j \neq i$;
		\item and the final equality comes from Bayes' rule.
	\end{enumerate*}
	Note that the approximation that $z_i$ is independent of $z_j$ where $j \in \{1,\dots,\nobs \},\; j \neq i$ is suboptimal as it does not capture the correct conditional distribution of $\z$.
	Since the inference scales the posterior density to integrate to 1, the approximation of $y_i$ being independent of $y_j$ plays no role in the RHF inference process as it only scales the inferred density. 

	Next, the hidden states are updated using linear regression: 
	\begin{equation}
		\label{eq:rhf-regression}
		\xxae_i = \xxe_i + \frac{\operatorname{cov}(x_i, z_j)}{\operatorname{cov}(z_j, z_j)} (\zzae_j - \zze_j),
	\end{equation}
	for all $j \in \{1, \dots, \nobs \}$, $i \in \{ 1, \dots, \nstate\}$, and $e \in \{ 1, \dots, \nens\}$ as in the RHF~\cite{Anderson_2010_Rhf}.
	Here, $\operatorname{cov}(v_1, v_2)$ denotes the empirical covariance of the random variables $v_1$ and $v_2$, following the local least squares framework for ensemble filtering~\cite{Anderson_2003_Local}. 

	It is also possible to formulate the univariate RHF posterior in terms of an optimal transport map from the prior to the posterior \cite[Theorem 2.9]{Santambrogio_2015_OT} as
	\begin{equation}
		\label{eq:opt-transport-rhf}
		\xxae_i = \prc_{x_i \mid y_i}^{-1} \circ \prc_{x_i} \left(\xxe_i \right) \quad \forall e \in \{ 1, \dots, \nens\}.
	\end{equation}
	This is similar to the MARHF~\cite{Anderson_2020_Marhf}, in the sense that the analysis particle ranks match the forecast (prior) particle ranks (unlike the MARHF, where the analysis particle ranks match the EnKF analysis ranks).
	%
	%
	
	\subsubsection{QCEFF}
	\label{sssec:QCEFF}
	
	In QCEFF~\cite{Anderson_2022_Qceff1,Anderson_2023_Qceff2,Anderson_2024_Qceff3}, one uses the covariance after Gaussian anamorphosis to perform linear regression on the unobserved variables.
	Formally, one first constructs the following probit variables:
	\begin{subequations}
		\begin{align}
			\tilde{x}^{[e]}_i &= \phi^{-1} \circ \prc_{x_i}(\xxe_i), \\
			\tilde{z}^{[e]}_i &= \phi^{-1} \circ \prc_{z_i}(\zze_i), \\
			\tilde{z}^{\mathrm{a}[e]}_i &= \phi^{-1} \circ \prc_{z_i}(\zzae_i), 
		\end{align}
	\end{subequations}
	where $\phi^{-1}(u) = \sqrt{2} \operatorname{erf}^{-1}(2u - 1)$ is the probit function.
	Next, one calculates the analysis using linear regression in the probit space, then maps it back to the physical space, as follows:
	\begin{align}
		\label{eq:qceff-regression}
		\tilde{x}^{\mathrm{a}[e]}_i &= \tilde{x}^{[e]}_i + \frac{\operatorname{cov}( \tilde{x}_i,  \tilde{z}_j)}{\operatorname{cov}( \tilde{z}_j, \tilde{z}_j)} (\tilde{z}^{\mathrm{a}[e]}_j - \tilde{z}^{[e]}_j), \\
		\xxae_i &= \prc^{-1}_{x_i} \circ \phi (\tilde{x}^{\mathrm{a}[e]}_i),
	\end{align}
	for all $j \in \{1, \dots, \nobs \}$, $i \in \{ 1, \dots, \nstate\}$, and $e \in \{ 1, \dots, \nens\}$.
	According to \cite{Grooms_2022_Nonlinear}, the QCEFF assumes a Gaussian copula for the regression.
	%
	
	\subsection{Copulas}
	\label{ssec:copula}
	
	We briefly review the theory and estimation of copulas.
	A copula represents a cumulative joint distribution in quantile space.
	Consider the random variable $\x = [x_1,\dots,x_n] \in \Rspace^n$. 
	The associated copula $\cpj$ is a function that maps the set of marginal quantiles $u_i = \prc_{x_i}(x_i)$, $i=1,\dots,n$, to the multivariate CDF of $\x$~\cite{Nelsen_2006_Copulas,Charpentier_2007_Copulas} as
	\begin{equation}
		\cpj : [0, 1]^n \to [0, 1], \quad 
		\cpj(\*{u}) = \prc_{\x}(\x), \quad \*{u} = [u_1,\dots,u_n], \quad  u_i = \prc_{x_i}(x_i),
	\end{equation}
	where $0 \leq u_i \leq 1$ for all $i \in \{1, \dots, n\}$.
	The copula has the following properties:
	\begin{enumerate}[label=(\roman*)]
		\item $\cpj(\*{u}) = 0$ if any $u_i = 0$.
		\item $\cpj(\*{u}) = u_i$ if $u_j = 1$ for all $j \in \{1, \dots, n\} \backslash\{ i\}$.
		\item $\cpj(\*{u})$ is non-decreasing in each argument, i.e., $\cpj(\*{u}) \leq \cpj(\tilde{\*{u}})$ whenever $u_j \leq \tilde{u}_j$ and $u_i = \tilde{u}_i$ for all $i \in \{1, \dots, n \} \backslash\{ j\}$.
	\end{enumerate}
	Sklar's theorem~\cite{Nelsen_2006_Copulas} states that every multivariate joint probability density function can be represented as a product of the marginals and the copula density function: 
	\begin{equation}
		\pr(\x) = \frac{\partial^{n}}{\partial x_1\dots\partial x_n} \prc_{\x}(\x) = \left( \frac{\partial^{n}}{\partial u_1  \dots \partial u_n} \cpj(\*{u}) \right) \prod_{i = 1}^{n} \frac{\partial u_i}{\partial x_i} = \cpd(\*{u}) \prod_{i = 1}^{n} \pr(x_i),
	\end{equation}
	where $\cpd(\*{u}) \coloneqq \frac{\partial^{n}}{\partial u_1  \dots \partial u_n} \cpj(\*{u}) : [0,1]^n \to \Rspace$ is the copula density function.
	
	Next, consider the conditional density $\pr(x_{j} \mid \x_{1 : j-1})$ where $1 < j \leq n$.
	Applying Sklar's theorem gives: 
	\begin{equation}
		\label{eq:cond-copula-density}
		\pr(x_{j} \mid \x_{1 : j-1}) = \frac{\pr( \x_{1:j})}{\pr( \x_{1 : j-1})} = \frac{\cpd(\*u_{1:j}) \prod_{i = 1}^{j} \pr(x_i)}{\cpd(\*u_{1:j - 1}) \prod_{i = 1}^{j - 1} \pr(x_i)} = \cpd(u_j \mid \*u_{1 : j -1}) \,\pr(x_j),
	\end{equation}
	where $\cpd(u_{j} \mid \*u_{1 : j - 1}) = \cpd(\*u_{1:j}) \slash \cpd(\*u_{1:j - 1})$ is the conditional copula density function.
	In the following sections, with some abuse of notation, we write copulas as functions of the random variable $\x$ rather than $\*{u}$:
	\begin{align}
		\cpd(\*u) = \cpd\Big(\prc_{x_1}(x_1), \prc_{x_2}(x_2), \dots, \prc_{x_n}(x_n)\Big) \eqqcolon \cpd(\x).
	\end{align}

	\section{The copula rank histogram filter}
	\label{sec:crhf}
	
	We now derive the copula rank histogram filter and discuss its implementation details.
	Consider again the system \cref{eq:model,eq:measurement,eq:observables} with $\x$ the hidden state variables, $\z$ the observable variables, and $\y$ the measurements; each component $y_i$ is a measurement of component $z_i$, $i=\{1,\dots,\nobs\}$. 
	The posterior of the state-observable system given by \cref{eq:bayes_cond} reads:
	\begin{equation}
		\label{eq:state-observable}
		\pr(\x, \z \mid \y) =  \pr(\z \mid \y)\,\pr(\x \mid \z).
	\end{equation}
	The first term $\pr(\z \mid \y)$ is expanded similar to \cref{eq:obs-independence}, without making any independence assumption as
	\begin{align}
		\label{eq:obs-cop-lik-cond}
		\pr(\z \mid \y) = \frac{\pr(y_1 \mid z_1)\, \pr(z_1)}{\pr(y_1)} \prod_{i = 2}^{\nobs} \frac{\pr(y_i \mid z_i)\, \pr(z_i \mid \z_{1:i-1})}{\pr(y_i \mid \y_{1:i-1})}.
	\end{align}
	As the constant denominators ($\pr(y_1), \; \pr(y_i \mid \y_{1:i-1})$) play no role in the univariate inference, we denote \cref{eq:obs-cop-lik-cond} as
	\begin{equation}
		\label{eq:obs-cop-lik-cond-ind}
		\pr(\z \mid \y) = \pr(z_1 \mid y_1) \prod_{i = 2}^{\nobs} \pr(z_i \mid y_i, \z_{1:i-1}),
	\end{equation}
	for brevity. Similarly, the term $\pr(\x \mid \z)$ is expanded using the chain rule as 
	\begin{equation}
		\label{eq:x-cond-z}
		\pr(\x \mid \z) = \pr(x_1 \mid \z) \prod_{i = 2}^{\nstate} \pr(x_i \mid \z, \x_{1:i-1}).
	\end{equation}
	\Cref{eq:state-observable,eq:obs-cop-lik-cond-ind,eq:x-cond-z} define a triangular filter (in the style of \cite{Spantini_2018_Coupling,Spantini_2022_Coupling}) as each variable depends only on the set of previous (defined lexicographically) variables i.e. $x_i$ depends on $\z$ and $\x_{1:i-1}$ only.
	The CoRHF methodology (summarized in \Cref{alg:corhf}), is described as follows: 
	\begin{enumerate}[label=(\roman*)]
		\item First, samples of the first observable $\zzae_1 \sim \pr(z_1 \mid y_1)$ for $e \in \{1, \dots, \nens\}$ are obtained using the reconstructed RHF posterior density.
		Sampling $\zzae_1 = \prc_{z_1 \mid y_1}^{-1}(\uue)$ can be done in three ways as described in \cref{eq:samp-sto,eq:samp-qsto,eq:samp-qdet}.
		Here, $\prc_{z_1 \mid y_1}^{-1} \circ \prc_{z_1}$ is a transport map from the prior probability to the posterior probability spaces. 
		\item For each observable $i$, in the order $i = 2,\dots, \nobs$, use the already obtained samples $\zae_{1:i-1}$ and \cref{eq:obs-cop-lik-cond-ind} to obtain posterior samples $\zzae_i$, $e \in \{1, \dots, \nens\}$ as follows:
		\begin{equation}
			\label{eq:pos-cop-lik}
			\zzae_i \sim \pr(z_i \mid y_i, \z_{1:i-1} = \zae_{1:i-1}) \propto \pr(y_i \mid z_i) \cpd(z_i \mid \z_{1:i-1} = \zae_{1:i-1}) \pr(z_i).
		\end{equation}
		This is realized by calling \Cref{alg:rhf-like} with input samples $\zze_i$, and $\pr(y_i \mid z_i) \cpd(z_i \mid \z_{1:i-1} = \zae_{1:i-1})$ as the scaling function, and sampling the resulting distribution with any method from \cref{eq:samp-sto,eq:samp-qsto,eq:samp-qdet}.
		The piecewise continuous approximation of the prior $\pr(z_i)$ is constructed similar to the rank histogram filter (see in \Cref{sssec:RHF}).
		The piecewise continuous reconstruction of the product of the observation likelihood and the conditional copula density $\pr(y_i \mid z_i) \cpd(z_i \mid \zae_{1:i-1})$ is done as described in \cref{eq:piecewise-g} from \Cref{alg:rhf-like}. 
		This results in sampling from an RHF-like posterior density, which multiplies the likelihood, prior density, and the conditional copula density. 
		\item Finally, the analysis hidden states $\xxae_i$  are obtained one by one in the order $i = \{1, \dots, \nstate \}$ by sampling the corresponding factor in \cref{eq:x-cond-z}:
		\begin{equation}
			\label{eq:pos-cop}
			\xxae_i \sim \pr\left(x_i \mid \z = \zae, \x_{1:i-1} = \xae_{1:i-1}\right) \propto \cpd\left(x_i \mid \z = \zae, \x_{1:i-1} = \xae_{1:i-1}\right)\,  \pr(x_i),
		\end{equation}
		across $i = \{1, \dots, \nstate \}$ and samples $e \in \{1, \dots,\nens \}$. 
		This is implemented by calling \Cref{alg:rhf-like} with $\xxe_i$ as the input samples, and $\cpd(x_i \mid \z = \zae, \x_{1:i-1} = \xae_{1:i-1})$ as the scaling function, and sampling the resulting distribution with any method from \cref{eq:samp-sto,eq:samp-qsto,eq:samp-qdet}.
	\end{enumerate}
	\begin{algorithm}
		\caption{CoRHF}
		\label{alg:corhf}
		\KwIn{Prior samples : $\ze$, $\xe$ for all $e \in \{1, \dots,\nens\}$.}
		Obtain $\prc_{z_1}$ using \Cref{alg:rhf-like} on $\zze_1$, $e \in \{1, \dots,\nens\}$ and $p(y_1 \mid z_1)$.\\
		Compute $\zzae_1$ according to any sampling method from \cref{eq:samp-sto,eq:samp-qsto,eq:samp-qdet}.\\
		\For{$i = 2$ to $\nobs$}{
			\ForAll{$e = 1$ to $\nens$}{
				Obtain $\prc_{z_i \mid y_i, \zae_{1:i-1}}^{[e]}$ using \cref{alg:rhf-like} on $z^{[1:\nens]}_i$ and $p(y_i \mid z_i) \cpd(z_i \mid \ze_{1:i-1} = \zae_{1:i-1})$.\\
				Compute $\zzae_i$ according to any sampling method from \cref{eq:samp-sto,eq:samp-qsto,eq:samp-qdet}.
		}}
		\For{$i = 1$ to $\nstate$}{
			\ForAll{$e = 1$ to $\nens$}{
				Obtain $\prc_{x_i \mid \zae, \xae_{1:i-1}}^{[e]}$ using \cref{alg:rhf-like} on $x^{[1:\nens]}_i$ and $\cpd(x_i \mid \ze_{1:\nobs} = \zae_{1:\nobs}, \xe_{1:\nobs} = \xae_{1:\nobs})$.\\
				Compute $\xxae_i$ according to any sampling method from \cref{eq:samp-sto,eq:samp-qsto,eq:samp-qdet}.}}
		\KwRet{$\xae$.}
	\end{algorithm}

	While we extend the rank histogram assumption on the prior density and a piecewise constant likelihood, any approximation to respective functions can be used (see \cite{Anderson_2022_Qceff1} for more choices).
	\begin{remark}[Variable reordering]
		Similar to triangular maps, the posterior of any univariate only depends on the previous univariate analyses in the lexicographic order.
		Note that the internal ordering of the observables $\z$ and states $\x$ can be independently modified.
		While the posterior density remains invariant to variable reordering, the samples will change.
		It could be worthwhile to investigate variable order to account for long-distance correlations \cite{Katzfuss_2021_Local}, or allow for parallelism \cite{Anderson_2007_Scalable}.
	\end{remark}

	\subsection{A scalar filtering example}

\begin{figure*}[!ht]
		\centering
		\includegraphics[width=\linewidth]{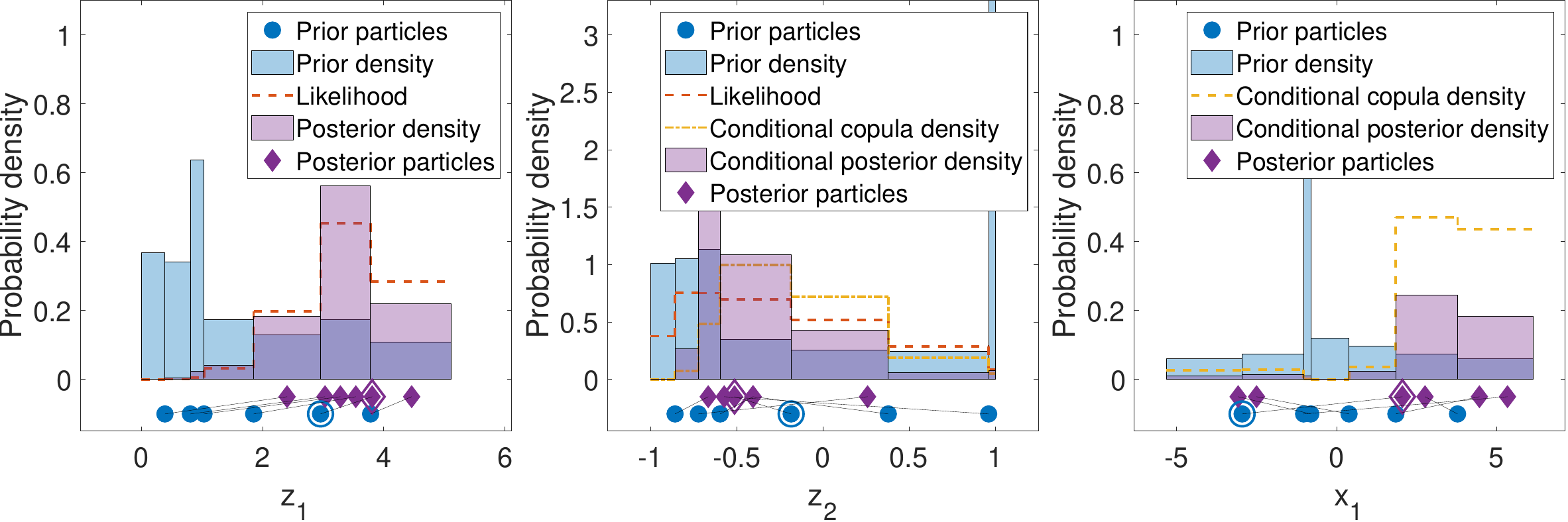}
		\caption{The conditional copula density and the conditional posterior are associated with the outlined particle in the middle figure. Similarly, the conditional posterior is associated with the outlined particle in the rightmost figure.}
		\label{fig:corhfs}
	\end{figure*}

	Consider a scalar random variable $\x \in \Rspace$, with samples shown as the prior particles in the right subplot of \Cref{fig:corhfs}.
	This, is non-linearly observed where $\Hn(x) = [
	|x| , \sin{x}
	]^\top$.
	The prior observables are shown in the left and middle subplots of \Cref{fig:corhfs} respectively.
	First, we assimilate on $z_1 = |x|$ using its likelihood as shown in the left subplot of \Cref{fig:corhfs}.
	Next, we assimilate on $z_2 = \sin(x)$, using its likelihood and the conditional copula density conditioned on $z_1$ as shown in the middle subplot of \Cref{fig:corhfs}.
	Finally, we sample $\xxa$ conditioned on $\za$ as shown in the right subplot of \Cref{fig:corhfs}.
	While we extend the rank histogram assumption on the prior density and a piecewise constant likelihood, any approximation to respective functions can be used (see \cite{Anderson_2022_Qceff1} for more choices).

	\subsection{Two-dimensional filtering examples}
	
\begin{figure}[p]    
		\centering
		\subfigure[Setup]{\includegraphics[width=0.4\linewidth]{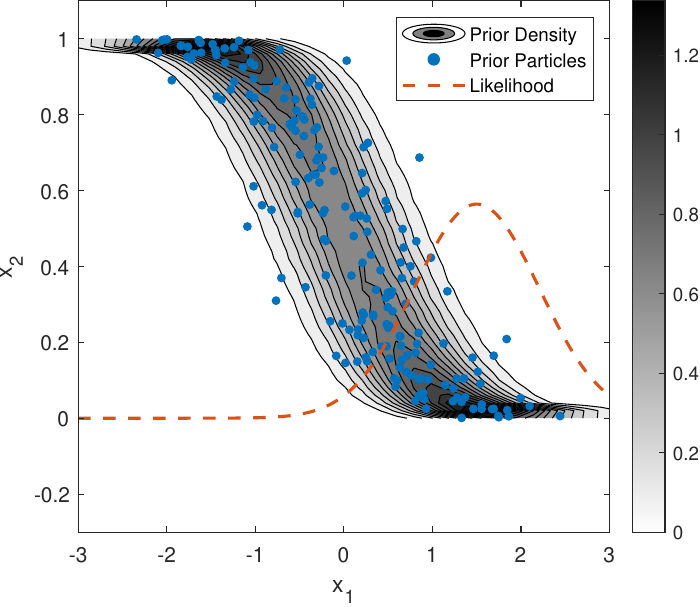}\label{fig:nb1}}
		\subfigure[ETKF analysis]{\includegraphics[width=0.4\linewidth]{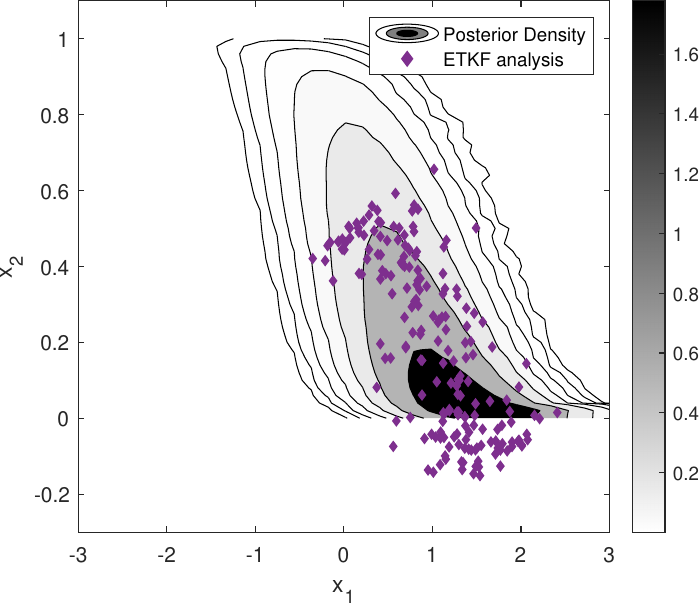}\label{fig:nb2}}
		\subfigure[ETPF analysis]{\includegraphics[width=0.4\linewidth]{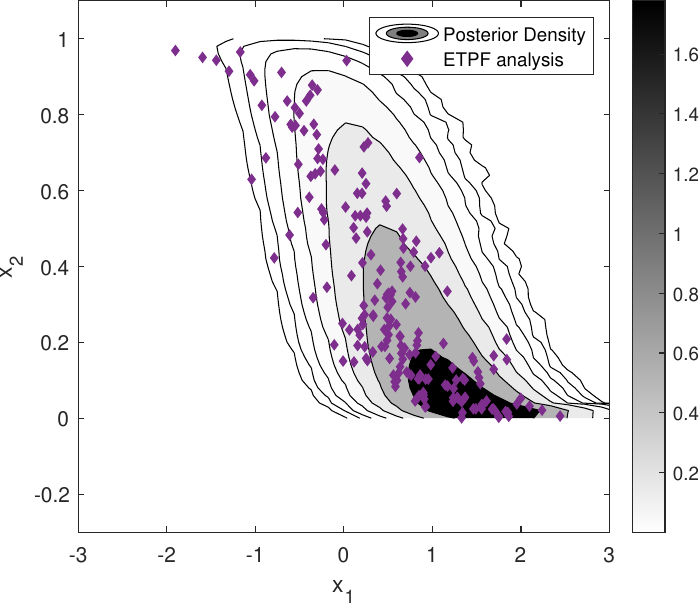}\label{fig:nb3}}
		\subfigure[RHF analysis]{\includegraphics[width=0.4\linewidth]{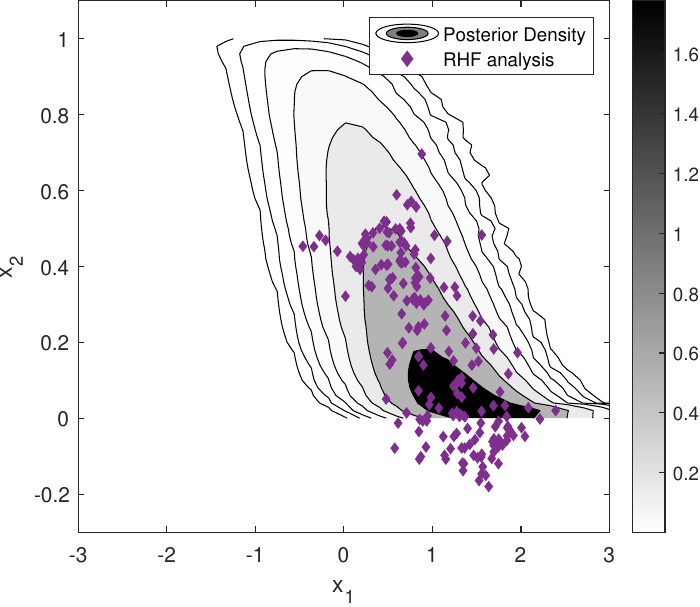}\label{fig:nb4}}
		\subfigure[QCEFF analysis]{\includegraphics[width=0.4\linewidth]{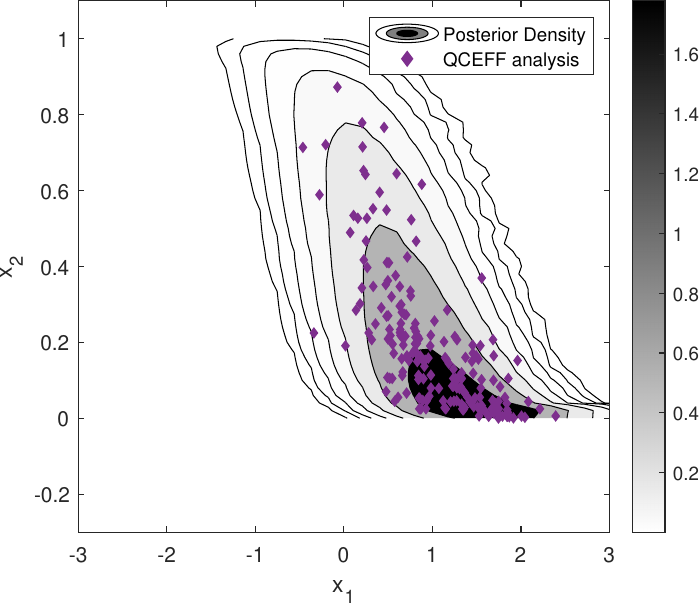}\label{fig:nb5}}
		\subfigure[CoRHF analysis]{\includegraphics[width=0.4\linewidth]{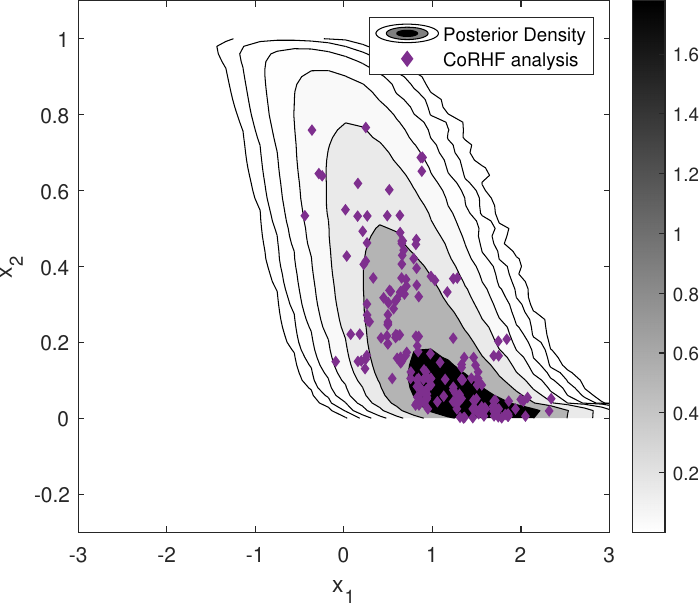}\label{fig:nb6}}
		\caption{Example demonstrating the analysis samples for multiple filters for a normal-beta prior density and a normal observation. Filters such as the ETKF and RHF, which use correlation information directly, end up with samples inconsistent with the posterior. The ETPF, QCEFF, and CoRHF all give samples consistent with the posterior.}
		\label{fig:nb-example}
	\end{figure}

	We now compare different filters on two simple two-dimensional densities where only the first variable is observed.
	In \Cref{fig:nb-example} (adapted from \cite{Anderson_2022_Qceff1}), we look at a normal beta density where the marginals in the $x_1$ and $x_2$ dimensions are normally and beta distributed, respectively. 
	We observe the $x_1$ variable ($z=x_1$) with a Gaussian observation error as shown in \Cref{fig:nb1}.
	The prior density is visualized by generating $10^7$ samples from the analytical prior, and computing normalized two-dimensional histogram counts like in \cite{Anderson_2022_Qceff1}.
	Then, the analytical posterior density is calculated by multiplying the analytical prior density and the observational likelihood as per Bayes' rule.
	We compare the sampled posteriors produced by several different filters: ETKF (\Cref{fig:nb2}), ETPF (\Cref{fig:nb3}), RHF (\Cref{fig:nb4}), QCEFF (\Cref{fig:nb5}), and the CoRHF (\Cref{fig:nb6}).
	The ETKF and RHF, both of which perform Gaussian regression, sample from under the x-axis, which is a zero probability region.
	However, ETPF, QCEFF, and CoRHF sample the posterior correctly.
	
	However, for a multimodal density as shown in the next figure~\Cref{fig:spiral-example}, only the ETPF (\Cref{fig:sp3}) and CoRHF (\Cref{fig:sp6}) sample the posterior.
	However, ETPF does not capture the spread as accurately as the CoRHF. 
	This could be resolved by adding stochastic noise, but this may not always work. 
	The other filters ETKF (\Cref{fig:sp2}), RHF (\Cref{fig:sp4}) and QCEFF (\Cref{fig:sp5}) fail to capture the posterior due to Gaussian assumptions in the estimation of $x_2$.
	\begin{figure}[p]    
		\centering
		\subfigure[Setup]{\includegraphics[width=0.4\linewidth]{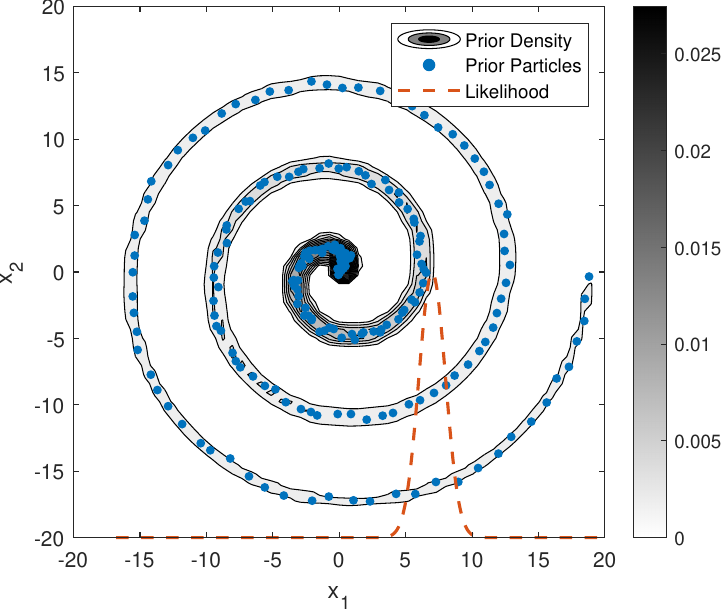}\label{fig:sp1}}
		\subfigure[ETKF analysis]{\includegraphics[width=0.4\linewidth]{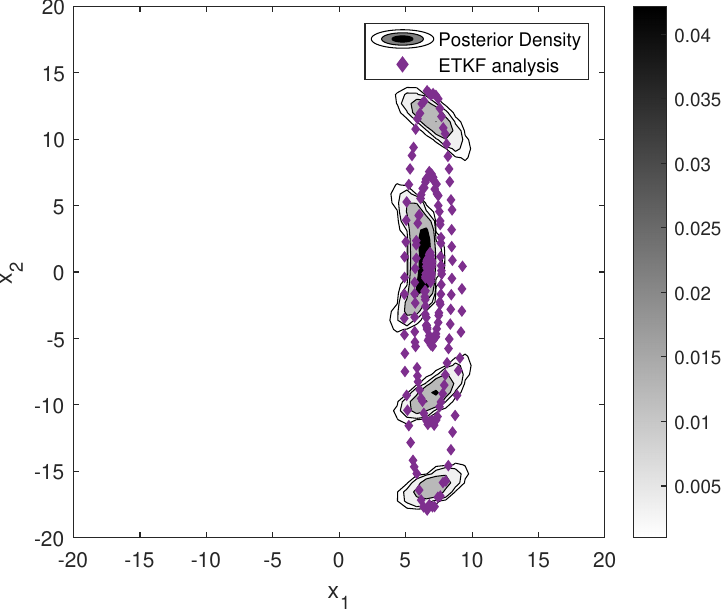}\label{fig:sp2}}
		\subfigure[ETPF analysis]{\includegraphics[width=0.4\linewidth]{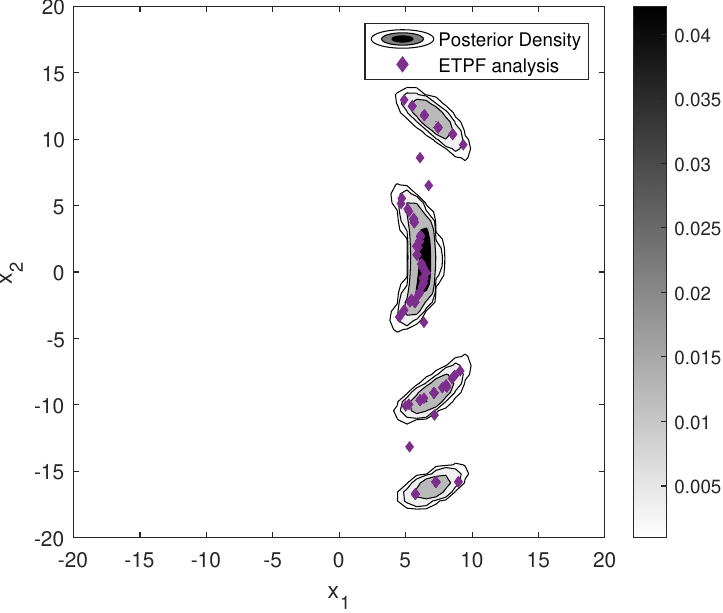}\label{fig:sp3}}
		\subfigure[RHF analysis]{\includegraphics[width=0.4\linewidth]{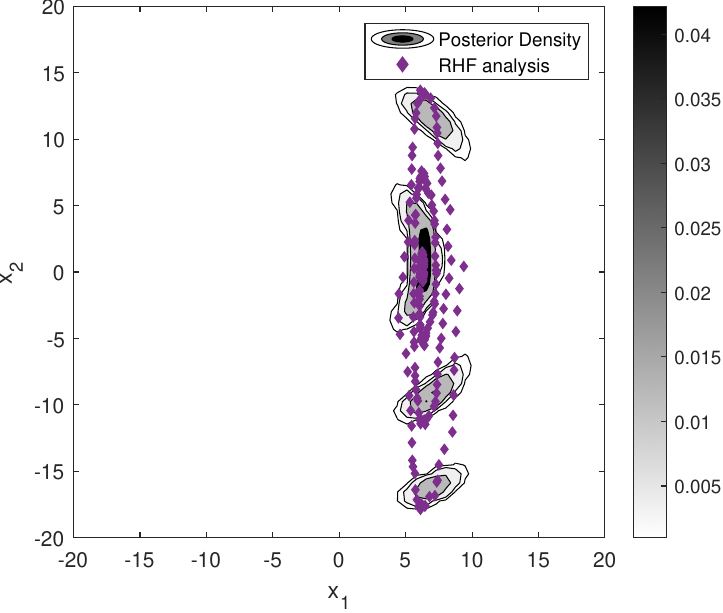}\label{fig:sp4}}
		\subfigure[QCEFF analysis]{\includegraphics[width=0.4\linewidth]{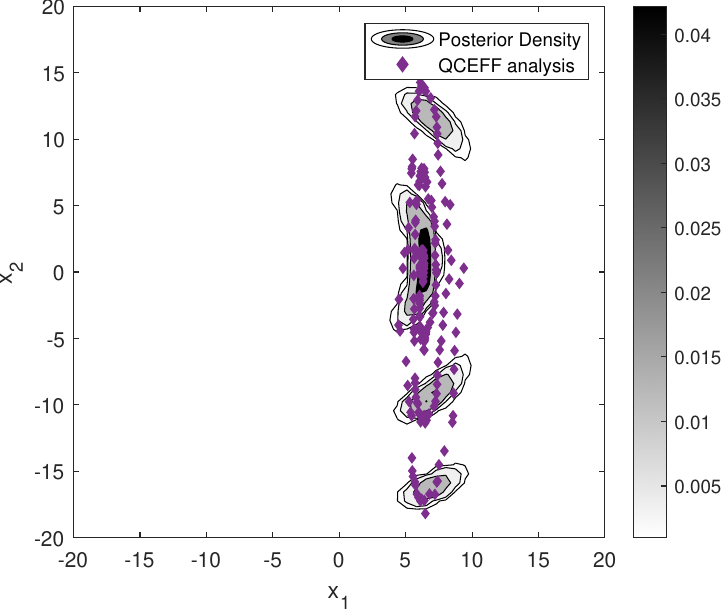}\label{fig:sp5}}
		\subfigure[CoRHF analysis]{\includegraphics[width=0.4\linewidth]{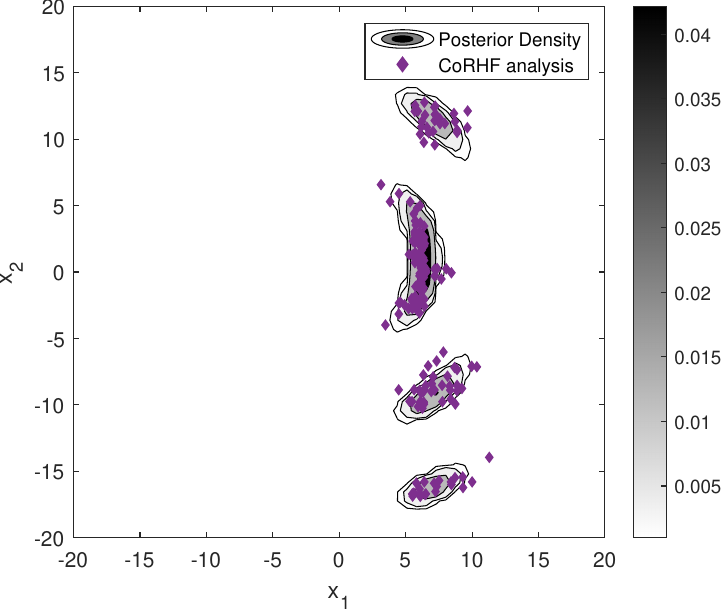}\label{fig:sp6}}
		\caption{Example demonstrating the analysis samples for multiple filters for a spiral prior density and a normal observation. Filters other than the ETPF and CoRHF fail to sample the multimodal posterior correctly. The CoRHF has a qualitatively better spread for each mode, unlike the ETPF, which samples along a line.}
		\label{fig:spiral-example}
	\end{figure}
	
	\section{Estimation of conditional copula densities}
	\label{sec:crhf-implementation}
	Efficient estimation of conditional copula densities is an essential ingredient of the CoRHF (discussed in \Cref{sec:crhf}).
	\subsection{Empirical representation of copula densities}
	Given a set of samples $\left\{ \ue \right\}_{e = 1}^{\nens}$, an empirical copula estimate is given as 
	\begin{equation}
		\cpj(\*{u}) = \frac{1}{\nens} \sum_{e = 1}^{\nens} \mathbbm{1}(u_1 \geq \uue_1, u_2 \geq \uue_2, \dots, u_{n} \geq \uue_{n}) = \frac{1}{\nens} \sum_{e = 1}^{\nens} \prod
		_{i = 1}^{n} \mathbbm{1}(u_i \geq \uue_i),
	\end{equation}
	where $\mathbbm{1}(u > \tilde{u})$ refers to the indicator function which returns $1$ if $u > \tilde{u}$ or $0$ otherwise. 
	Here, $\*{u}$ can refer to any combination of $\prc_{z_i}(z_i)$ and $\prc_{x_i}(x_i)$ variables as required for the conditional copula density computation from \cref{eq:pos-cop-lik,eq:pos-cop}.
	The indicator function (or the Heaviside step function) is non-smooth due to the jump discontinuity, and its derivative cannot be calculated. 
	To allow for differentiability, a smooth approximation to the scalar indicator function is made by a scalar sigmoid function represented by $\sigma$ as: 
	\begin{align}
		\cpj(\*{u}) &= \frac{1}{\nens} \sum_{e = 1}^{\nens} \prod
		_{i = 1}^{n} \sigma(u_i; \uue_i), \\
		\label{eq:copula-density-estimate}
		\cpd(\*{u}) &= \frac{1}{\nens} \sum_{e = 1}^{\nens} \prod
		_{i = 1}^{n} \frac{\partial \sigma(u_i; \uue_i)}{\partial u_i} = \frac{1}{\nens} \sum_{e = 1}^{\nens} \prod
		_{i = 1}^{n} \sigma'(u_i; \uue_i).
	\end{align}
	The function $\sigma'$ is estimated using kernel density estimation.
	The kernel choice must be supported on the domain $[0, 1]$, and be $0$ everywhere else.
	There are many possible choices, such as truncated Gaussian kernels: 
	\begin{equation}
		\label{eq:kernel-trunc-gaussian}
		\sigma'(u; u^{[e]}) = \left\{ \begin{matrix}
			\frac{1}{h\sqrt{2 \pi}}\exp \left(-\frac{(u - u^{[e]})^2}{2h^2} \right) & 0 \leq u \leq 1,\\
			0 & \mathrm{otherwise,}
		\end{matrix}  \right.
	\end{equation}
	beta kernels:
	\begin{equation}
		\label{eq:kernel-beta}
		\sigma'(u; u^{[e]}) = \mathfrak{B}\left(u^{[e]}; \frac{u}{h} + 1, \frac{1 - u}{h} + 1 \right),
	\end{equation}
	or boundary-corrected beta kernels~\cite{Chen_1999_Beta}:
	\begin{equation}
		\label{eq:kernel-beta-bc}
		\sigma'(u; u^{[e]}) = \left\{ \begin{array}{ll}
			\mathfrak{B}\left(u^{[e]}; \frac{u}{h}, \frac{1 - u}{h} \right) & 2h \leq u \leq 1 - 2h,\\
			\mathfrak{B}\left(u^{[e]}; \rho(u, h), \frac{1 - u}{h} \right) & 0 \leq u < 2h, \\
			\mathfrak{B}\left(u^{[e]}; \frac{u}{h}, \rho(1 - u, h) \right) & 1 - 2h < u \leq 1,
		\end{array}  \right.
	\end{equation}
	where $\rho(u, h) = 2h^2 + 2.5 -\sqrt{4h^4 + 6h^2 + 2.25 - u^2 - \frac{u}{h}}$~\cite{Chen_1999_Beta} and $\mathfrak{B}(u; \alpha, \beta) = \frac{u^{\alpha - 1}(1 - u)^{\beta - 1}}{B(\alpha, \beta)}$ is the beta distribution. All above kernels are parametrized by a bandwidth parameter $h$ to control the spread. 
	
	It is the authors' experience that the boundary-corrected beta kernel (\cref{eq:kernel-beta-bc}) performs slightly better than the beta kernel, both of which performed significantly better than the truncated Gaussian.
	This is likely because of the high amount of boundary bias in the truncated Gaussian.
	For these reasons, the boundary-corrected beta kernel is chosen for all the experiments.
	
	The kernel bandwidth $h$ is a free parameter that needs to be estimated based on data. 
	As in \cite{Patilea_2009_Beta}, one may choose a bandwidth that minimizes the estimator's mean integrated squared error.
	However, we take the simpler approach from \cite{Renault_2004_Beta}, and choose a scaled standard deviation (represented by $\operatorname{stddev}(\cdot)$) of the samples for the bandwidth as
	\begin{equation}
		h = \alpha \cdot \operatorname{stddev}(u_i^{[1:\nens]}) \cdot \nens^{-\frac{2}{5}}.
	\end{equation}
	Since CoRHF uses a rank histogram prior, a constant bandwidth can be used for each state as $u_i^{[1:\nens]} = \left\{ \frac{j}{\nens + 1} \right\}_{j = 1}^{\nens}$. 
	Note that $\displaystyle \lim_{\nens \to \infty} \sigma(u_i^{[1:\nens]}) \to 12^{-\frac{1}{2}}$, which is the standard deviation of the standard uniform distribution as $u_i^{[1:\nens]}$ are samples of the standard uniform density.  
	The smoothness of the kernel density estimate is controlled with $\alpha$, which scales the spread of the samples. 
	In this work, $\alpha$ is treated as a problem-specific tunable hyperparameter.

	\begin{figure*}[!ht]    
		\centering
		\includegraphics[width=\linewidth]{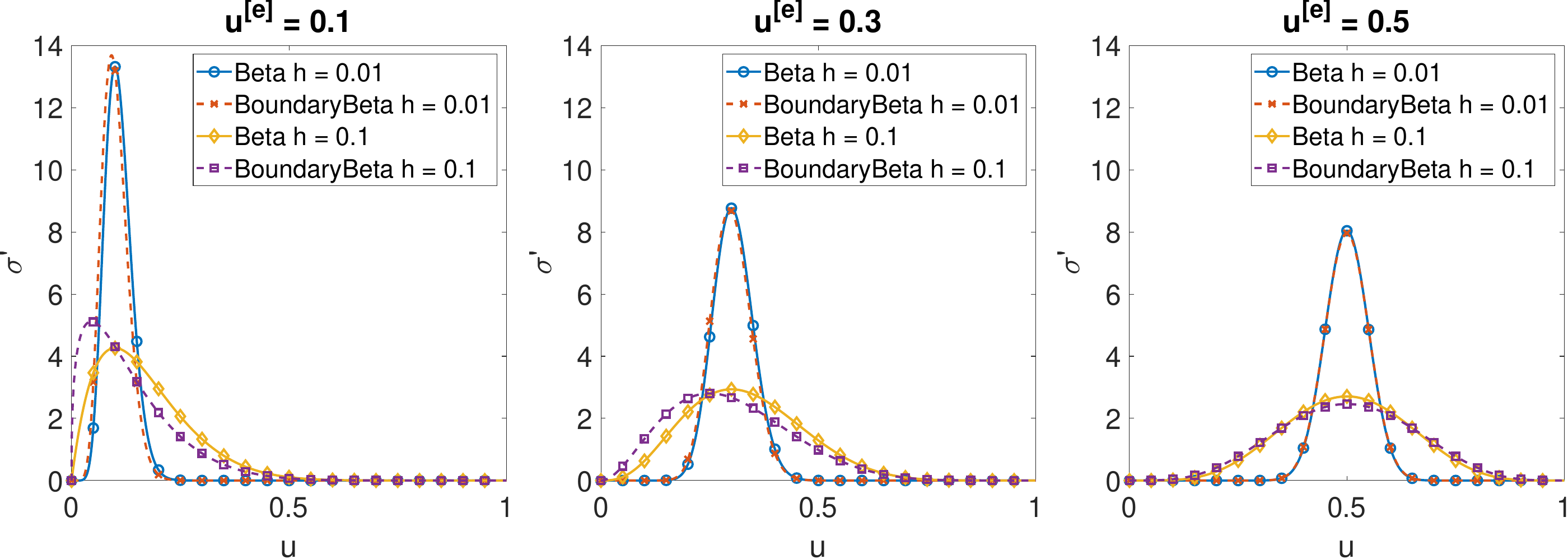}
		\caption{Kernel density estimates of $\sigma'(u; \uue)$ with the beta and boundary corrected beta kernels at varying values of $u^{[e]}$ and bandwidth $h$.}
		\label{fig:kern-example}
	\end{figure*}

	\begin{example}
		\Cref{fig:kern-example} shows the beta kernel and boundary corrected beta kernel $\sigma'(u; \uue)$ for $\uue = 0.1, 0.3, 0.5$ and bandwidths $h = 0.1, 0.01$.
		The two kernels show similar behavior in region $2h \leq u \leq 1 - 2h$, but show differing behavior near the boundaries.
	\end{example}
	
	\subsection{Efficient estimation of conditional copula densities}
	Three main elements aid the efficient estimation of conditional copula densities for CoRHF:
	\begin{enumerate}
		\item The constant denominator $\cpd(\*u_{1 : j -1} = \*u^{\mathrm{a}[e]}_{1 : j-1})$ of the conditional copula density 
		\begin{equation}
			\label{eq:copula-density-estimate-actual}
			\cpd(u_j \mid \*u_{1 : j -1} = \*u^{\mathrm{a}[e]}_{1 : j-1}) = \frac{\cpd( \*u_{1 : j-1} = \*u^{\mathrm{a}[e]}_{1 : j-1},u_j)}{\cpd(\*u_{1 : j -1} = \*u^{\mathrm{a}[e]}_{1 : j-1}) },
		\end{equation}
		is not computed. 
		It is irrelevant to posterior computation in \cref{eq:pos-cop-lik,eq:pos-cop} as the respective posterior in each case is rescaled to ensure a total area of 1.
		\item Next, the most efficient way to estimate the numerator of \cref{eq:copula-density-estimate-actual} is to first substitute $\*u_{1 : j-1} = \*u^{\mathrm{a}[e]}_{1 : j-1}$ into \cref{eq:copula-density-estimate}. 
		This gives
		\begin{align}
			\label{eq:cop-comp}
			\cpd(u_j, \*u_{1 : j-1} = \*u^{\mathrm{a}[e]}_{1 : j-1}) &= \frac{1}{\nens} \sum_{\varepsilon = 1}^{\nens} \sigma'(u_j; u^{[\varepsilon]}_j)
			\left( \prod_{i = 1}^{j-1} \sigma'(u_i = u^{\mathrm{a}[e]}_i; u^{[\varepsilon]}_i) \right) \\
			\nonumber
			&= \frac{1}{\nens} \sum_{\varepsilon = 1}^{\nens} \sigma'(u_j; u^{[\varepsilon]}_j) \cdot \gamma_j^{[\varepsilon]},
		\end{align}
		where the constants $\gamma_j^{[\varepsilon]}$ are first estimated to give a univariate interpolant $\cpd(u_j, \*u_{1 : j-1} = \*u^{\mathrm{a}[e]}_{1 : j-1})$.
		Note that this must be done individually for each $e \in \{1, \dots, \nens\}$.
		It is possible to construct a $j$-dimensional interpolation hypersurface as 
		\begin{align}
			\label{eq:copula-wrong}
			\widetilde{\cpd}(\*u_{1 : j}) &= \frac{1}{\nens} \sum_{\varepsilon = 1}^{\nens} 
			\left( \prod_{i = 1}^{j} \sigma'(u_i; u^{[\varepsilon]}_i) \right),
		\end{align}
		and then substitute $\*u_{1 : j-1} = \*u^{\mathrm{a}[e]}_{1 : j-1}$ into $\widetilde{\cpd}(\*u_{1 : j})$ to obtain $\cpd(u_j, \*u_{1 : j-1} = \*u^{\mathrm{a}[e]}_{1 : j-1})$.  
		But this is not recommended due to the prohibitive cost.
		\item Estimating \cref{eq:kernel-beta,eq:kernel-beta-bc} requires computing the beta function which requires the gamma function as $B(\alpha, \beta) = \frac{\Gamma(\alpha)\Gamma(\beta)}{\Gamma(\alpha + \beta)}$.
		Direct estimation of gamma functions for extreme values can result in numerical overflows, see \cite[Chapter 6.1]{Press_2007_NumericalRecipes} for more details.
		Calculating the log-beta function $\log B(\alpha, \beta) = \log\Gamma(\alpha) + \log\Gamma(\beta) - \log\Gamma(\alpha + \beta)$ with the help of log-gamma functions avoids the overflow problem.
		Additionally, we exploit this opportunity to transform the product in the computation of $\gamma_j^{[\varepsilon]}$ into a sum i.e. given as
		\begin{equation}
			\gamma_j^{[\varepsilon]} = \prod_{i = 1}^{j-1} \sigma'(u_i = u^{\mathrm{a}[e]}_i; u^{[\varepsilon]}_i) = \exp \left( \sum_{i = 1}^{j-1} \log \sigma'(u_i = u^{\mathrm{a}[e]}_i; u^{[\varepsilon]}_i)  \right).
		\end{equation}
		Now, computing $\log\sigma'$ for both beta kernels (\cref{eq:kernel-beta}) and boundary-corrected beta kernels (\cref{eq:kernel-beta-bc}) requires computing the logarithm of the beta distribution which is $\log\mathfrak{B}(u; \alpha, \beta) = (\alpha - 1)\log u + (\beta - 1)\log(1 - u) - \log B(\alpha, \beta)$.
	\end{enumerate}

	\subsubsection{Localization}
	
	For high dimensional problems, it is infeasible to estimate a conditional copula density, while suffering from sample deficit.
	For this purpose, we taper the dependency structure on the conditional copula density as follows:
	\begin{equation}
		\begin{split}
			\gamma_j^{[e]} &= \prod_{i = 1}^{j-1} \sigma'(u_i = u^{\mathrm{a}[e]}_i; \uue_i)^{\rho(d(u_{j}, u_{i}))}\\&= \exp \left( \sum_{i = 1}^{j-1} \rho(d(u_{j}, u_{i})) \log \left( \sigma'(u_i = u^{\mathrm{a}[e]}_i; \uue_i) \right) \right),    
		\end{split}
	\end{equation}
	where $d(u_{j}, u_{i})$ models some notion of distance between the two random variables $u_{j}$ and $u_{i}$, and $\rho$ is a tapering function.
	For example, if both $u_j$ and $u_i$ are marginal uniforms of state variables ($u_j = \prc_{x_j}(x_j)$ and $u_i = \prc_{x_i}(x_i)$) that live on a geospatial grid, then $d(u_{j}, u_{i})$ could be the physical distance between the corresponding grid points between the states $x_j$ and $x_i$.
	Ultimately, the distance function $d$ controls the influence of variable $z_i$ or $x_i$ on the variable $z_j$ or $x_j$ in the copula, where a larger distance implies a smaller influence and vice versa.
	In this work, we use the Gaspari-Cohn function, where $\zeta = d(u_{j}, u_{i})$~\cite{Gaspari_1999_GCfun,Hamill_2001_Localization}. 
	\begin{equation}
		\label{eq:gc-loc}
		\rho(\zeta) = \left \{ \begin{matrix}
			&-\frac{1}{4}\zeta^5 + \frac{1}{2}\zeta^4 + \frac{5}{8}\zeta^3 -\frac{5}{3}\zeta^2 + 1 &\text{ if } 0 \leq \zeta < 1, \\ &\frac{1}{12}\zeta^5 - \frac{1}{2}\zeta^4 + \frac{5}{8}\zeta^3 +\frac{5}{3}\zeta^2 - 5\zeta + 4 &\text{ if } 1 \leq \zeta < 2, \\
			&0 &\text{ otherwise. }
		\end{matrix} \right.
	\end{equation}
	%
	
	\section{Experiments}
	\label{sec:expt}
	
	We test our formulation on two problems: the Lorenz '63~\cite{Lorenz_1963_L63} and Lorenz '96~\cite{Lorenz_1996_L96}.
	The methodology is evaluated on the spatiotemporal root mean squared error for both the forecast and analysis states.
	The analysis RMSE is given by 
	\begin{equation}\label{eq:RMSE}
		\mathrm{RMSE} = \sqrt{\frac{1}{\nk\nstate}\sum_{k = 1}^{\nk} \left\| \xtrk - \mu(\xak) \right\|^2_2}
	\end{equation}
	where $\xtrk$ is the latent true state at time $t^k$ (see \cref{eq:measurement}) and $\mu(\xak)$ is the mean of the analysis particles at $t^k$.
	The forecast RMSE is given by replacing $\xak$ with the forecast $\xk$.
	We compare the CoRHF, with the EnKF~\cite{Evensen_1994_EnKF}, RHF~\cite{Anderson_2010_Rhf}, QCEFF~\cite{Anderson_2022_Qceff1}, and the Ensemble Gaussian Mixture filter~\cite{Popov_2024_EnGMF} (EnGMF).
	RHF, QCEFF, and CoRHF all use the same assumptions when building the prior density.
	RHF and QCEFF use the optimal transport quantile deterministic sampling in \cref{eq:samp-qdet,eq:opt-transport-rhf}, whereas CoRHF uses quantile stochastic sampling without replacement as in \cref{eq:samp-qsto}. 
	Specifically, the latter filters assume a rank histogram prior between the samples, with a flat tail (details described in the specific experiment).
	All methods except the EnGMF perform stochastic inflation of the forecast ensemble of observations as 
	\begin{equation}
		\widetilde{\z}^{[e]} = \z^{[e]} + \obserr^{[e]}, \text{ where } \obserr^{[e]} \sim \pro,
	\end{equation}
	for use in the respective filter.
	This is theoretically necessary for only the EnKF.
	However, this results in much lower RMSE for the RHF, QCEFF and CoRHF in the following experimental settings. 
	
	\subsection{Lorenz '63}
	\label{sec:expt-l63}
	
	The three-variable Lorenz '63 problem~\cite{Lorenz_1963_L63} with the canonical chaotic parameters is described by the ordinary differential equation system:
	\begin{equation}\label{eq:lorenz63}
		\begin{split}
			\dot{x} &= \sigma(y - x), \\ \dot{y} &= x(\rho - z) - y, \quad \quad \begin{pmatrix}
				\sigma & \rho & \beta
			\end{pmatrix} = \begin{pmatrix}
				10 & 28 & 8/3
			\end{pmatrix}, \\ \dot{z} &= xy - \beta z.
		\end{split}
	\end{equation}

	\paragraph*{Observations.}
	New observations are made available every $\Delta t = 0.5$, resulting in highly non-linear dynamics and non-Gaussian priors.
	We observe the two-norm distance from the equilibrium point $ \*p = \begin{bmatrix}
		\sqrt{\beta(\rho - 1)} & \sqrt{\beta(\rho - 1)} & (\rho - 1)
	\end{bmatrix}^\top$ with a positive half-Gaussian observation error (with scale parameter $\*R = 1$). 
	Formally, the observation operator is 
	\begin{equation}
		\Hn(\x) = (\x - \*p)^\top(\x - \*p), \quad \Hn : \Rspace^{3} \to \Rspace,
	\end{equation}
	and the observation error is given by $\obserr = |\widehat{\obserr}|$ where $\widehat{\obserr} \sim \operatorname{Normal}(0, \*R)$.
	This particular choice for the observation error distribution is made for two main reasons.
	\begin{enumerate}[label=(\roman*)]
		\item As the observation is a distance, it cannot be negative if the noise is too large. 
		\item It is more challenging for traditional filters than the two-sided Gaussian distribution. 
		We prefer a challenging setting for this nonlinear, non-Gaussian filter as the filter performs comparably to traditional filters for linear-Gaussian settings.
	\end{enumerate}
	Note that while the observation error is drawn from a positive half-Gaussian distribution, the inflated ensemble observation $\widetilde{\z}^{[e]}$ could be spread on either side of the observation $\y$, requiring a full Gaussian likelihood (with covariance $\*R$ and mean zero).
	
	\paragraph*{Experiment setup.}
	For all filters, assimilation is done for 5500 assimilation steps, discarding the statistics from the first 500 steps as the spinup.
	To ensure the validity and robustness of the results, the experiments are repeated for 36 different observation trajectories. 
	The baseline RMSE for the problem is obtained from the bootstrap particle filter with $\nens = 10,000$.
	%
	%
	The prior densities for RHF, QCEFF, and CoRHF all use a flat tail.
	The tail length is $\min(2\operatorname{stddev}(x_i), 2(1.2)^\ell\operatorname{stddev}(x_i))$ where $\ell$ is chosen to be the smallest integer such that the perturbed observation $\tilde{y}_i$ lies inside the tail.
	\begin{figure*}[!ht]    
		\centering
		\includegraphics[width=0.8\linewidth]{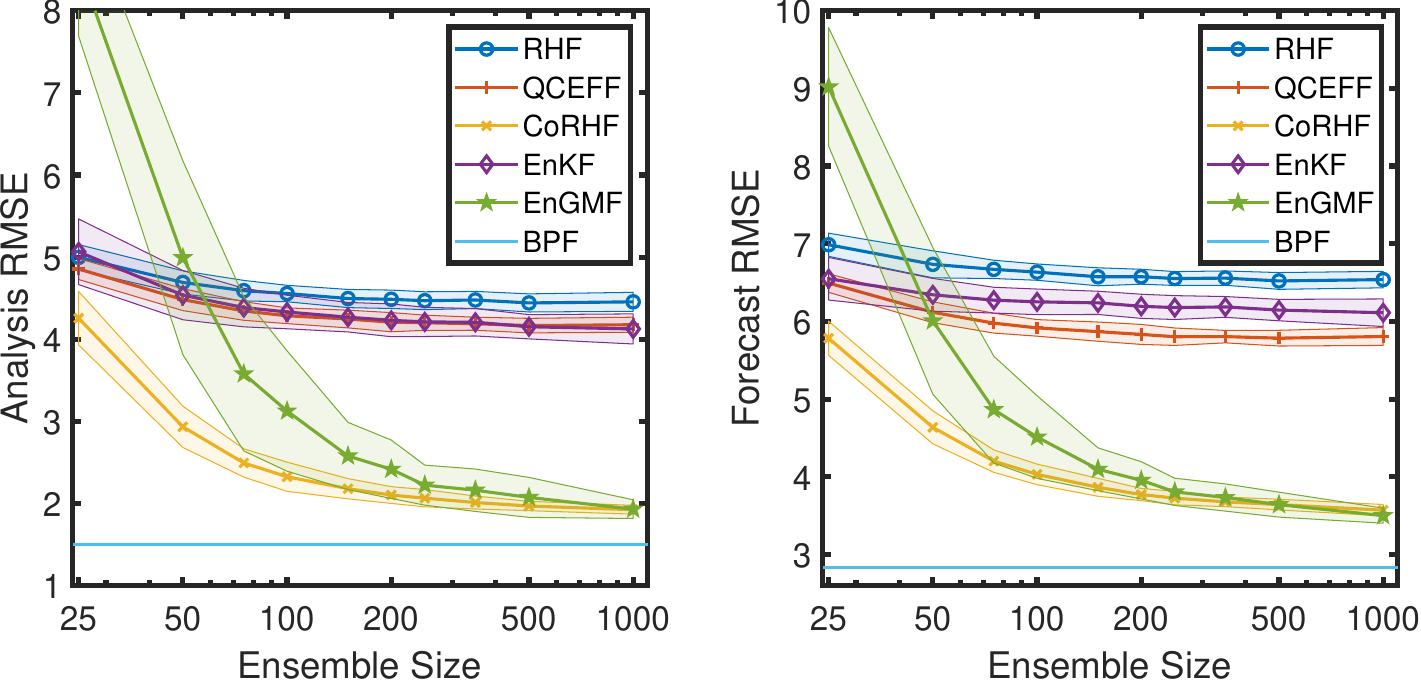}
		\caption{Mean RMSE (along with 2 standard deviations as shared) across 36 experiments for the Lorenz '63 problem. Note that the baseline bootstrap particle filter uses $10,000$ particles.}
		\label{fig:l63res}
	\end{figure*}

	\paragraph*{Results.}
	The analysis RMSE (left subplot of \Cref{fig:l63res}) and forecast RMSE (right subplot of \Cref{fig:l63res}) show similar trends.
	The CoRHF method shows better performance than the EnKF, RHF, and QCEFF counterparts and gets close to the bootstrap particle filter as $\nens$ becomes larger.
	As expected, the EnKF, RHF, and QCEFF show similar performance due to the suboptimality of correlation-based regression when representing the conditional relationship between the different variables (both observation and state).
	Multiple assumptions could be made on the prior densities and tails for the RHF, QCEFF, and CoRHF, which may provide different results.
	As $\nens$ becomes larger, EnGMF starts to perform comparably with CoRHF.
	
	\subsection{Lorenz '96} 
	\label{sec:expt-l96}
	Next, we run experiments on the Lorenz '96 problem~\cite{Lorenz_1996_L96} described by
	\begin{equation}\label{eq:lorenz96}
		\dot{x}_i = (x_{i+1} - x_{i-2})x_{i-1} - x_i + 8,
		\quad
		i \in \{1, \dots,{40}\}, \quad
		x_{-1} = x_{39},\quad x_{41} = x_{1}.
	\end{equation}
	\paragraph*{Observations.}
	Observations are available every $\Delta t = 0.2$.
	The absolute value of every alternate state is observed, i.e., $\Hn(\x) = [|\x|_1,|\x|_3,\dots,|\x|_{39}]$.
	The observation error is independent and half-Cauchy distributed, i.e., $\obserr = |\widehat{\obserr}|$ where every element $\widehat{\obserr}_i \sim \operatorname{Cauchy}(0, \gamma = 0.1)$ for $i \in \{1,2,\dots,20 \}$.
	As in the previous experiment, the inflated ensemble observation $\widetilde{\z}^{[e]}$ can lie on either side of the observation $\y$, requiring a full Cauchy distribution with $\gamma = 0.1$ in the likelihood computation.
	This setup results in non-linear priors with heavy-tailed (noisy) observations.
	
	\paragraph*{Experiment setup.}
	
	Assimilation is done for 2200 assimilation steps, discarding the statistics from the first 200 steps as the spinup.
	The experiments are run for 36 different observation trajectories to ensure robustness.
	The prior densities for RHF, QCEFF, and CoRHF all use a flat tail whose length is $2\operatorname{stddev}(x_i)$.
	A fixed scheme for the tail length was chosen since Cauchy observations could be extremely noisy (rendering the observation non-informative).
	We use \cref{eq:gc-loc} with the distance function being the cyclic distance between the state variables.
	Since the observations are just a subset of states, the distance is computed with the state (position) information. 

	\begin{figure*}[!ht]    
		\centering
		\includegraphics[width=0.8\linewidth]{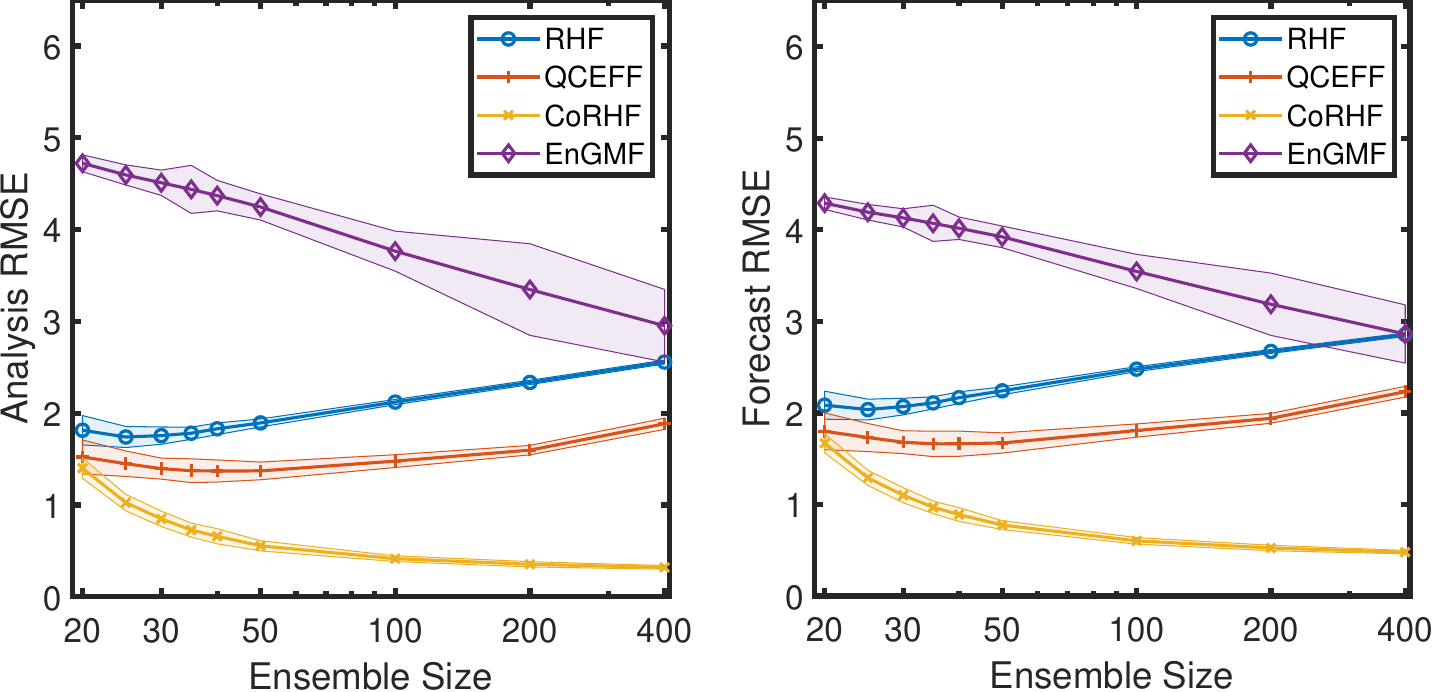}
		\caption{Mean RMSE (along with 2 standard deviations as shared) across 36 experiments for the Lorenz '96 problem.}
		\label{fig:l96res}
	\end{figure*}
	\paragraph*{Results.}
	Here, CoRHF does much better than RHF, QCEFF, and EnGMF on both the analysis RMSE (left subplot of \Cref{fig:l96res}) and forecast RMSE (right subplot of \Cref{fig:l96res}).
	CoRHF does markedly better because it accounts for the conditional dependence structure among the states.
	As seen in \cref{fig:l96res}, the RMSE for RHF and QCEFF increases with ensemble size.
	We believe that this is due to tuning, where the tail length was tuned for situations of $\nens < \nstate$, which is suboptimal for higher $\nens$.
	
	\section{Conclusions}
	\label{sec:conc}
	
	We successfully developed a copula rank histogram filter, called the CoRHF, for data assimilation. CoRHF is a triangular transportation filter that uses previously computed state analyses to construct conditional copula densities that inform the analysis of subsequent states. 
	This formulation is necessary because the existing triangular filters do not account for the conditional dependency structure between the observables and the states.
	Accounting for this is necessary to obtain the correct analysis particles in non-Gaussian, multimodal scenarios. 
	We also develop a computationally efficient methodology to estimate univariate conditional copula densities, including localization for high-dimensional systems. 
	We demonstrate the correctness of this filter on standard test problems, including the Lorenz '63 and Lorenz '96. 
	While more expensive than the RHF and QCEFF, CoRHF can capture the posterior more accurately due of including conditional information. 
	While we extend only the rank histogram reconstruction of prior density, one can use any approximate prior discussed in QCEFF~\cite{Anderson_2022_Qceff1}. 
	
	In future work, we plan to test CoRHF on more realistic and larger test problems involving environmental simulations.
	This would require variable reordering using ideas from \cite{Anderson_2007_Scalable} to allow for parallel computation in the univariate state inference. 
	This would also require investigating multiple different prior density assumptions, especially for the tails.
	The importance of tail choice has been noted in \cite{Simon_2012_Anamorphosis,Anderson_2010_Rhf}.

	
	\acknowledgements
	
	This work was supported by the Department of Energy, the National Science Foundation via awards DMS--2411069 and DMS--2436357, and by the Computational Science Laboratory at Virginia Tech.
	
	\section*{Data availability statement}
	Data sharing is not applicable to this work. Code for the filter can be made available on request.
	
	
	
	
	
	
		
		
	
	
	
	\bibliographystyle{IJ4UQStyle}
	\bibliography{references}
	
\end{document}